\documentclass[useAMS,usenatbib]{mn2e}

\usepackage[]{natbib}
\usepackage{graphicx}

\title[Transit timing of TrES-2]{Transit timing of TrES-2: a combined analysis of ground- and space-based photometry\thanks{Based on observations collected at the Centro Astron\'{o}mico Hispano Alem\'{a}n (CAHA) at Calar Alto, operated jointly by the Max-Planck Institut f\"{u}r Astronomie and the Instituto de Astrofísica de Andaluc\'{i}a (CSIC). Based on observations obtained with telescopes of the University Observatory Jena, which is operated by the Astrophysical Institute of the Friedrich-Schiller-University Jena. This paper includes data collected by the Kepler mission. Funding for the Kepler mission is provided by the NASA Science Mission directorate. This research has made use of the NASA Exoplanet Archive, which is operated by the California Institute of Technology, under contract with the National Aeronautics and Space Administration under the Exoplanet Exploration Program.}}
\author[St. Raetz et al.]{St. Raetz$^{1,2}$\thanks{E-mail:sraetz@cosmos.esa.int}, G. Maciejewski$^{3}$, Ch. Ginski$^{1}$, M. Mugrauer$^{1}$, A. Berndt$^{1}$, T. Eisenbeiss$^{1}$, \newauthor Ch. Adam$^{1}$, M. Raetz$^{4}$, T. Roell$^{1}$, M. Seeliger$^{1}$, C. Marka$^{1}$, M. Va\v{n}ko$^{5}$,  \L{}. Bukowiecki$^{3}$, \newauthor R. Errmann$^{1,6}$, M. Kitze$^{1}$, J. Ohlert$^{7,8}$, T. Pribulla$^{5}$, J.G. Schmidt$^{1}$,  D. Sebastian$^{9}$, \newauthor D. Puchalski$^{3}$, N. Tetzlaff$^{1}$, M.M. Hohle$^{1}$, T.O.B. Schmidt$^{1}$ and R. Neuh\"{a}user$^{1}$\\
$^{1}$Astrophysikalisches Institut und Universit\"{a}ts-Sternwarte, Schillerg\"{a}\ss{}chen 2-3, 07745 Jena, Germany\\
$^{2}$European Space Agency, ESTEC, SRE-S, Keplerlaan 1, 2201 AZ Noordwijk, The Netherlands\\
$^{3}$Centre for Astronomy, Faculty of Physics, Astronomy and Informatics, Nicolaus Copernicus University, Grudziadzka 5, 87-100 Torun, Poland \\
$^{4}$Private observatory Raetz, Stiller Berg 6, 98587 Herges-Hallenberg, Germany \\
$^{5}$Astronomical Institute, Slovak Academy of Sciences, 059 60 Tatransk\'{a} Lomnica, Slovakia \\
$^{6}$Abbe Center of Photonics, Friedrich-Schiller-Universit\"{a}t Jena, Max-Wien-Platz 1, 07743 Jena, Germany \\
$^{7}$Michael-Adrian-Observatorium, Astronomie-Stiftung Trebur, Fichtenstrasse 7, D-65468 Trebur, Germany \\
$^{8}$University of Applied Sciences, Technische Hochschule Mittelhessen, 61169 Friedberg, Germany \\
$^{9}$Th\"{u}ringer Landessternwarte Tautenburg, Sternwarte 5, D-07778 Tautenburg, Germany 
}
             
\begin{document}

\date{Accepted 2014 July 24.  Received 2014 July 24; in original form 2013 July 23}

\pagerange{\pageref{firstpage}--\pageref{lastpage}} \pubyear{2002}

\maketitle

\label{firstpage}

\begin{abstract}
Homogeneous observations and careful analysis of transit light curves can lead to the identification of transit timing variations (TTVs). TrES-2 is one of few exoplanets, which offer the matchless possibility to combine long-term ground-based observations with continuous satellite data. Our research aimed at the search for TTVs that would be indicative of perturbations from additional bodies in the system. We also wanted to refine the system parameters and the orbital elements. We obtained 44 ground-based light curves of 31 individual transit events of TrES-2. Eight 0.2\,--\,2.2-m telescopes located at six observatories in Germany, Poland and Spain were used. In addition, we analysed 18 quarters (Q0--Q17) of observational data from NASA's space telescope \textit{Kepler} including 435 individual transit events and 11 publicly available ground-based light curves. Assuming different limb darkening (LD) laws we performed an analysis for all light curves and redetermined the parameters of the system. We also carried out a joint analysis of the ground- and space-based data. The long observation period of seven years (2007--2013) allowed a very precise redetermination of the transit ephemeris. For a total of 490 transit light curves of TrES-2, the time of transit mid-point was determined. The transit times support neither variations on long time-scale nor on short time-scales. The nearly continuous observations of Kepler show no statistically significant increase or decrease in the orbital inclination $i$ and the transit duration $D$. Only the transit depth shows a slight increase which could be an indication of an increasing stellar activity. In general, system parameters obtained by us were found to be in agreement with previous studies but are the most precise values to date.
\end{abstract}

\begin{keywords}
planets and satellites: individual: TrES-2 -- stars: individual: GSC 03549-02811 -- planetary systems.
\end{keywords}

\section{Introduction}

A transit of TrES-2 was first observed in the summer of 2005 and published by \citet{2006ApJ...651L..61O}. With a period of $\sim$\,2.47\,d, TrES-2 is a hot Jupiter which shows a strong grazing eclipse. To determine the stellar and planetary properties of the TrES-2 system (see Table~\ref{Werte_TrES2}) photometric and spectroscopic follow-up observations were done by \citet{2007ApJ...664.1185H}, \citet{2007ApJ...664.1190S}, and \citet{2009A&A...507..523A}. \citet{2008ApJ...682.1283W} measured the Rossiter-McLaughlin effect and found a good alignment of the stellar spin and the planetary orbit which implies that TrES-2 orbits on a prograde orbit around its host star.\\ TrES-2 is in the field of view of the \textit{Kepler} space telescope, making it one of the photometrically best studied transiting exoplanets.\\ There have been controversial discussions about TrES-2.\citet{2009A&A...500L..45M} analysed two observed transits together with the three transits given in \citet{2007ApJ...664.1185H} and argued that the transit duration has shortened since 2006 by about $\sim$\,3.16 min, indicating a decrease of the orbital inclination. \citet{2009A&A...508.1011R} do not confirm this trend from the observation of five transits over a period of two years. New observations of two transit events by \citet{2010A&A...510A.107M} showed again a decrease of the inclination. But the analysis of one additional transit by \citet{2010ApJ...714..462S} showed only a non-significant change of the inclination. In \citet{2011EPJWC..1105007R} we also could not find any evidence for Transit Duration Variation. Additional nine transits of the NASA EPOXI mission of \citet{2011ApJ...726...94C} weakened the claims of the decreasing inclination further. The analysis of data from the \textit{Kepler} space telescope by \citet{2010ApJ...713L.160G} and \citet{2011ApJ...733...36K} (quarters 0 and 1) finally ruled out a change of the inclination as a function of time on the level predicted by \citet{2009A&A...500L..45M}. However, by modelling the transits of four observation quarters of \textit{Kepler} \citet{2012A&A...539A..97S} found a marginally significant positive slope of $\Delta i=(8\pm2)\times10^{-5}$ $^{\circ}$ per cycle which could be a hint for systematic variations.\\ \citet{2009A&A...498..567D} detected a companion candidate with a projected separation of $1.089\,\pm\,0.009$ arcsec to the TrES-2 host star in high-resolution images obtained with the \textit{AstraLux} Lucky Imaging camera of the Calar Alto Observatory. The companion candidate exhibits a magnitude difference in the $z'$ band of 3.43\,mag with respect to the TrES-2 host star. Although it was proven by detailed blend analysis that the TrES-2 host star is the source of the transit signal \citep{2006ApJ...651L..61O}, the companion candidate contaminates the photometric data, which leads to systematic errors in the light curve analysis. Since the published transit data were carried out with instruments that are not able to resolve the companion candidate, all light curves are contaminated by the so-called third light.  \citet{2013MNRAS.428..182B} re-observed the companion candidate to do a common proper motion analysis. Since the proper motion of the TrES-2 host star is too small for this purpose it is still unclear if the companion candidate is gravitationally bound.\\ The primary and also the secondary transit are well measured in different wavelength ranges. The observations in the infrared with Spitzer \citep{2010ApJ...710.1551O}, in the near-infrared with ground-based telescopes \citep{2010ApJ...717.1084C} and in the optical domain with \textit{Kepler} \citep{2011ApJ...733...36K,2011MNRAS.417L..88K}, yielded the albedo and hence provide constraints on the surface of the planet such as the dayside brightness temperature, the efficiency of dayside to nightside redistribution of heat, the atmospheric temperature structure (including a putative thermal inversion), the optical opacity but also the chemical composition of its atmosphere.\\ We observed TrES-2 at the University Observatory Jena since 2007. The analysis of 10 transits has already been published in \citet{2009AN....330..459R}, and of additional 12 transits in \citet{2011EPJWC..1105007R}. In this paper we describe the analysis of these 22 and additional nine transits that were observed between fall 2010 and spring 2013 at four different observatories. Furthermore, additional 435 transits from 18 quarters of \textit{Kepler} data were analysed in the same manner. The results yield refined system parameters and provide new limits on possible variations of them and the transit timing.\\ This paper is structured as follows. In \S2 we describe the observation, data reduction and photometry for our ground-based transit data. In \S3 we explain the fitting and analysis of the light curves. \S4 deals with the analysis of the space-based data including the determination of very precise system parameters and the investigation of the effect of limb darkening (LD). In \S5 we discuss a putative variation of the system parameters. Our results for the analysis of the transit timing are presented in \S6 and finally, in \S7, we conclude.

\begin{table}
\centering
\caption{System parameters of TrES-2 summarized from literature.}
\label{Werte_TrES2}
\begin{tabular}{cr@{\,$\pm$\,}lc}
\hline \hline
Parameter & \multicolumn{2}{c}{Value} & Ref \\ \hline
Epoch zero transit time $T_{0}$  [d] & \multicolumn{2}{c}{2453957.63492} &  [1] \\
 & & 0.00013 &  [1] \\
Orbital period $P$  [d] & 2.470614 & 0.000001 & [1] \\
Semi-major axis $a$ [au] & 0.03556 & 0.00075 &  [2] \\
Inclination $i$  [$^{\circ}$] & 83.62 & 0.14 & [2]\\
Eccentricity $e$ & \multicolumn{2}{c}{0} & [3]\\
Mass star $M_{\mathrm{A}}$  [M$_{\odot}$] & 0.98 & 0.06 & [4]\\
Radius star $R_{\mathrm{A}}$  [R$_{\odot}$] & 1.00 & 0.04 & [4]\\
Effective temperature $T_{\mathrm{eff}}$   [K] & 5795 & 73 & [5]\\
Surface gravity star log$\,g_{\mathrm{A}}$ & 4.457 & 0.004 & [6] \\
Metallicity $\left[ \frac{Fe}{H}\right]$ & 0.06 & 0.08 & [5]\\
Mass planet $M_{\mathrm{b}}$  [M$_{\mathrm{Jup}}$] & 1.26 & 0.05 & [6]\\
Radius planet $R_{\mathrm{b}}$  [R$_{\mathrm{Jup}}$] & 1.169 & 0.034 & [7] \\
Distance $d$ [pc] & 220 & 10 & [8] \\ 
Spectral type &  \multicolumn{2}{c}{G0V} & [4] \\ 
\hline \hline
\end{tabular}
\\ References:
[1] \citet{2009AN....330..459R}, [2] \citet{2009A&A...498..567D}, [3] \citet{2010ApJ...710.1551O}, [4] \citet{2006ApJ...651L..61O}, [5] \citet{2009A&A...507..523A}, [6] \citet{2010MNRAS.408.1689S}, [7] \citet{2011ApJ...726...94C}, and [8] \citet{2007ApJ...664.1190S}
\end{table}

\section{Observation, data reduction and photometry}

In addition to our previously published observations \citep{2009AN....330..459R,2011EPJWC..1105007R}, we measured nine transits at four observatories between 2010 September and 2013 April. Altogether we collected 44 (30 complete and 14 partial) light curves  of 31 individual transit events. We used eight 0.20\,--\,2.2-m telescopes located at six observatories in Germany, Poland and Spain. The participating observatories with their telescopes and instruments are summarized in Table~\ref{CCD_Kameras}. A summary of all observations is given in Table~\ref{Beobachtungslog_TrES2}. \\ The data reduction and photometry were performed uniformly for all observations. The photometric data were reduced following standard procedures including substraction of bias (as overscan, only if available) and dark, and division of a sky flat-field. We calibrated the CCD images using the \begin{scriptsize}IRAF\end{scriptsize}\footnote{\begin{scriptsize}IRAF\end{scriptsize} is distributed by the National Optical Astronomy Observatories, which are operated by the Association of Universities for Research in Astronomy, Inc., under cooperative agreement with the National Science Foundation.} routines \textit{darkcombine}, \textit{flatcombine} and \textit{ccdproc}.\\ Aperture photometry was performed with a dedicated \begin{scriptsize}IRAF\end{scriptsize} task \textit{chphot} which is based on the standard \begin{scriptsize}IRAF\end{scriptsize} routine \textit{phot} but allows us to do a simultaneous photometry of all field stars in a single image. The position of each star on every image was determined using \begin{scriptsize}ECLIPSE\end{scriptsize} Jitter \citep{1997Msngr..87...19D}. Differential magnitudes were calculated using the method of the optimized artificial comparison star developed by \citet{2005AN....326..134B}. The algorithm uses as many stars as possible (all available field stars) and calculates an artificial comparison star by weighting the stars according to their magnitude errors and variability. The final light curve is produced by  comparison of the TrES-2 host star with this artificial standard star. \\ For the photometry 10 different aperture radii were tested while the annulus for sky subtraction was kept fixed \citep[in our previous publications][we only used one fixed aperture radius therefore a re-analysis was necessary.] {2009AN....330..459R,2011EPJWC..1105007R}. The aperture that produced light curves with the best precision i.e. the lowest data point scatter was finally chosen. \\ We also tested a varying annulus for sky subtraction (different starting radius and width) while the aperture were kept fixed and the optimized artificial comparison star were calculated out of the same stars. Since the aperture and the artificial comparison star are optimized for a certain annulus no better precision could be achieved during this test. But also the calculation of a new artificial comparison star optimized for a different annulus yielded no better results. Thus fixing the annulus for sky subtraction with a subsequent optimization of aperture radius and artificial comparison star seems reasonable in our case.

\begin{table*}
\caption{Observatories and instruments which observed transits of TrES-2.}
\label{CCD_Kameras}
\begin{tabular}{cccccccc}
\hline \hline
Observatory & Long. (E) & Lat. (N) & Elevation & Telescope $\diameter$ & Camera & No. of Pixels & Pixel scale \\ 
& [$^{\circ}$] & [$^{\circ}$] & [m] & [m] & & & [arcsec per pixel] \\ \hline
University Observatory Jena & 11.48 & 50.93 & 370 & 0.25 & CTK$^{a}$ & 1024\,x\,1024 & 2.23 \\
& & & & 0.60 & STK$^{b}$ & 2048\,x\,2048 & 1.55  \\
& & & & 0.25 & CTK-II & 1056\,x\,1027 & 1.19  \\
& & & & 0.20 & RTK & 765\,x\,510 & 0.63  \\
Wendelstein Observatory & 12.01 & 47.70 & 1838 & 0.80 & MONICA$^{c}$  & 1024\,x\,1024 & 0.50 \\
Herges--Hallenberg & 10.55 & 50.69 & 431 & 0.20 & ST6 & 375\,x\,242 & 2.18 \\
& & & & 0.20 & G2-1600 & 1536\,x\,1024 & 1.88 \\
Michael Adrian Observatory & 8.41 & 49.93 & 103 & 1.20 & STL-6K\,3 & 1536\,x\,1024* & 0.40 \\
Calar Alto Observatory & 357.45 & 37.22 & 2168 & 2.20 & CAFOS$^{d}$ & 2048\,x\,2048 & 0.47 \\
Toru\'{n} Centre for Astronomy & 18.58 & 52.98 & 87 & 0.60 & STL-1001 & 512x512* & 1.39 \\
\hline \hline
\end{tabular}
\\
$^{\ast}$ $2 \times 2$ binning mode. \\
$^{a}$ \citet{2009AN....330..419M}, $^{b}$ \citet{2010AN....331..449M}, $^{c}$ \citet{1990ASPC....8..380R}, $^{d}$ \citet{1994S&W....33..516M}

\end{table*}

\subsection{University Observatory Jena photometry}

We observed TrES-2 for a total of 27 nights at the University Observatory Jena from 2007 March to 2012 June. Because of simultaneous observations with up to three telescopes (60\,cm Schmidt, 25 cm Cassegrain and 20 cm refractor on the same mount) we collected 36 light curves. Due to weather conditions and technical problems in some of the nights the data quality varies. Only 19 of these 27 transit epochs are fully covered. The remaining transits could be observed only partially or show gaps in the data due to passing clouds. Observations were performed in Bessell $V$, $R$ or $I$ filters \citep{1990PASP..102.1181B}. The exposure times varied from 15\,s to 50\,s with the 60\,cm Schmidt telescope, from 60\,s to 200\,s with the 25\,cm Cassegrain telescope and from 80\,s to 300\,s with the 20\,cm refractor, depending on weather conditions, airmass, and telescope focusing.

\subsection{Additional photometry}

As already described in \citet{2009AN....330..459R} we obtained two transit observations at the Wendelstein Observatory and one additional amateur observation from 2007 July to September. We re-analysed these observations.\\ From 2009 August to 2011 August we collected another four light curves from three observatories. Parameters of the observatories can be found in Table~\ref{CCD_Kameras} and are explained in detail in the following sections. The data were reduced and analysed in the same way as described above.

\subsubsection{Amateur observations}

Two complete transit light curves (2009 August 15th and 2009 August 30th) were obtained  with an 8-inch Schmidt--Cassegrain telescope (f/D\,=\,10) located in Herges--Hallenberg, Germany. With a G2-1600 CCD camera from Moravian Instruments Inc. with 1536\,$\times$\,1024 pixels and a pixel scale of 1.88 arcsec per pixel, we could observe a field of view of $48\,\times\,32\,$\,arcmin$^{2}$. The exposure times of the white-light observations (without filter) were 120\,s. 

\subsubsection{Michael Adrian Observatory}

The Trebur 1-m telescope (T1T) at the Michael Adrian Observatory in Trebur, Germany, was used to observe one complete transit on 2010 October 24th. The 1.2-m Cassegrain telescope is equipped with a SBIG STL6303 CCD camera. The 1536\,$\times$\,1024 pixels with a pixel scale of 0.4$''/pixel$ in 2\,$\times$\,2 binning mode correspond to a field of view of $10.2\times6.8$\,arcmin$^{2}$. The Bessell $R$ band observations with an exposure time of 15\,s were done under excellent weather conditions. 

\subsubsection{Calar Alto 2.2-m telescope}

On 2011 August 9th one transit of TrES-2 was observed as back-up of project H11-2.2-011 on the Calar Alto 2.2-m telescope.\\ For the observations we used the Calar Alto Faint Object Spectrograph (CAFOS) in imaging mode and 2\,$\times$\,2 binning. We windowed the field of view to $7.5\times4.5$\,arcmin$^{2}$ to shorten the read-out time. The instrument was heavily defocused (PSF of stars showed 'donut'-like structure) to allow for an exposure time of 30\,s. Observations were performed in the Cousins $R$ band. 

\subsubsection{Toru\'n Centre for Astronomy}

The 0.6-m Cassegrain telescope at the Centre for Astronomy of the Nicolaus Copernicus University in Toru\'n (Poland) was used to observe a transit on 2013 March 05. A $1024\times1024$-pixel SBIG STL-1001E CCD camera was used as a detector, giving a field of view of $11.8 \times 11.8$ \,arcmin$^{2}$ with a scale of $0.69$ arcsec per pixel. Observations were carried out without any filter to increase the instrument efficiency. We applied a $2 \times 2$ binning mode to shorten the readout time. The telescope was guided manually to keep star positions on the CCD matrix within several pixels. The exposure time was 10\,s (12\,s for a few images at the beginning of the run). Observations began 10\,min before the expected start of the transit, and continued an hour after the end of the event. The sky was clear during the run.

\begin{table*}
\caption{Summary of the observations: $N_{\mathrm{exp}}$ -- Number of exposures, $T_{\mathrm{exp}}$ -- exposure times.}
\label{Beobachtungslog_TrES2}
\begin{tabular}{lccccccc}
\hline \hline
Date & Epoch$^{a}$ & Camera$^{b}$ & Filter & $N_{\mathrm{exp}}$ & $T_{\mathrm{exp}}$ [s] & Average data point & rms [mmag] \\ 
& & & & &  & cadence  [s] &  \\ \hline
13 Mar. 2007 & 87 & CTK & $I$ &147 & 60 & 90 & 6.1 \\
03 May 2007 & 108 & CTK & \textit{I} &135 & 60 & 90 & 6.7 \\
17 Jul. 2007 & 138 & CTK* & \textit{I} &70 & 60 & 90 & 4.8 \\
26 Jul. 2007 & 142 & MONICA & \textit{R} & 157 & 30 & 112 & 1.9 \\
16 Sept. 2007 & 163 & CTK & \textit{I} &180 & 60 & 92 & 4.5 \\
& & MONICA & \textit{R} & 137 & 30 & 130 & 2.8 \\
& & ST6 & Clear & 180 & 60 & 105 & 11.9 \\
21 Sept. 2007 & 165 & CTK* & \textit{I} &217 & 60 & 93 & 4.5 \\
13 Oct. 2007 & 174 & CTK & \textit{I} &149 & 60 & 93 & 5.3 \\
26 Jun. 2008 & 278 & CTK & \textit{I} &145 & 60 & 89 & 3.7 \\
28 Sept. 2008 & 316 & CTK & \textit{I} &158 & 60 & 89 & 5.7 \\
03 Oct. 2008 & 318 & CTK* & \textit{I} &48 & 60 & 89 & 3.6 \\
11 Apr. 2009 & 395 & CTK & \textit{V} & 172 & 60 & 89 & 6.1 \\
21 Apr. 2009 & 399 & CTK & \textit{I} &218 & 60 & 89 & 5.8\\
28 May 2009 & 414 & CTK & \textit{I} &175 & 60 & 89 & 4.0 \\ 
02 Jun. 2009 & 416 & CTK & \textit{I} &137 & 60 & 89 & 4.7 \\
15 Aug. 2009 & 446 & STK & \textit{V} & 397 & 30, 25 & 39 & 2.7 \\
& & RTK & \textit{V} & 183 & 80 & 85 & 8.5 \\
& & G2-1600 & Clear & 113 & 120 & 126 & 3.1 \\
20 Aug. 2009 & 448 & STK* & \textit{R} & 128 & 30, 25 & 38 & 14.9 \\
& & CTK* & \textit{V} & 23 & 180 & 209 & 4.9 \\
& & RTK* & \textit{I} &16 & 300 & 301 & 4.2 \\
30 Aug. 2009 & 452 & G2-1600 & Clear & 61 & 120 & 126 & 2.6 \\
26 Sept. 2009 & 463 & STK & \textit{V} & 249 & 35 & 48 & 7.0 \\
& & RTK & \textit{V} & 50 & 200\,s & 210 & 10.7 \\
27 Nov. 2009 & 488 & CTK & \textit{R} & 137 & 60, 80 & 104 & 6.9 \\
24 Apr. 2010 & 548 & CTK & \textit{V} & 52 & 200 & 229 & 4.2 \\
10 Jun. 2010 & 567 & CTK & \textit{R} & 72 & 120 & 149 & 3.1 \\
27 Jul. 2010 & 586 & STK* & \textit{R} & 245 & 35, 30 & 42 & 3.8 \\
24 Oct. 2010 & 622 & STK & \textit{R} & 410 & 20, 18, 15 & 30 & 4.2 \\
& & CTK-II & \textit{R} & 126 & 90 & 92 & 3.1 \\
& & STL-6K\,3 & \textit{R} & 563 & 20,15 & 72$^{c}$ & 1.7 \\
29 Oct. 2010 & 624 & STK* & \textit{R} & 196 & 25 & 36 & 1.7 \\
22 Mar. 2011 & 682 & STK & \textit{R} & 242 & 35 & 49 & 2.8 \\
& & CTK-II* & \textit{V} & 86 & 120 & 123 & 3.0\\
27 Mar. 2011 & 684 & STK* & \textit{R} & 228 & 30 & 43 & 2.6 \\
& & CTK-II* & \textit{V} & 150 & 60 & 63 & 5.2 \\
04 Aug. 2011 & 737 & STK* & \textit{R} & 223 & 25 & 39 & 3.1 \\
& & CTK-II* & \textit{V} & 89 & 90 & 93 & 3.3 \\
09 Aug. 2011 & 739 & CAFOS& \textit{R} & 195 & 35 & 60 & 1.4 \\
25 Sept. 2011 & 758 & STK* & \textit{R} & 80 & 50 & 64 & 1.5 \\
& & CTK-II & \textit{R} & 165 & 80 & 83 & 2.7 \\
29 May 2012 & 858 & CTK-II & \textit{R} & 199 & 60 & 63 & 4.8 \\
05 Mar 2013 & 971 & STL-1001 & Clear & 813 & 10, 12 & 68$^{c}$ & 1.7 \\
\hline \hline
\end{tabular}
\\
$^{\ast}$ partial transit \\
$^{a}$ calculated using the ephemeris in \citet{2009AN....330..459R}\\
$^{b}$ for a description see Table~\ref{CCD_Kameras}\\
$^{c}$ average cadence for the final binned light curve
\end{table*}

\section{Light curve analysis}
\label{lc_analysis}

\begin{figure}
  \centering
  \includegraphics[width=0.16\textwidth,angle=270]{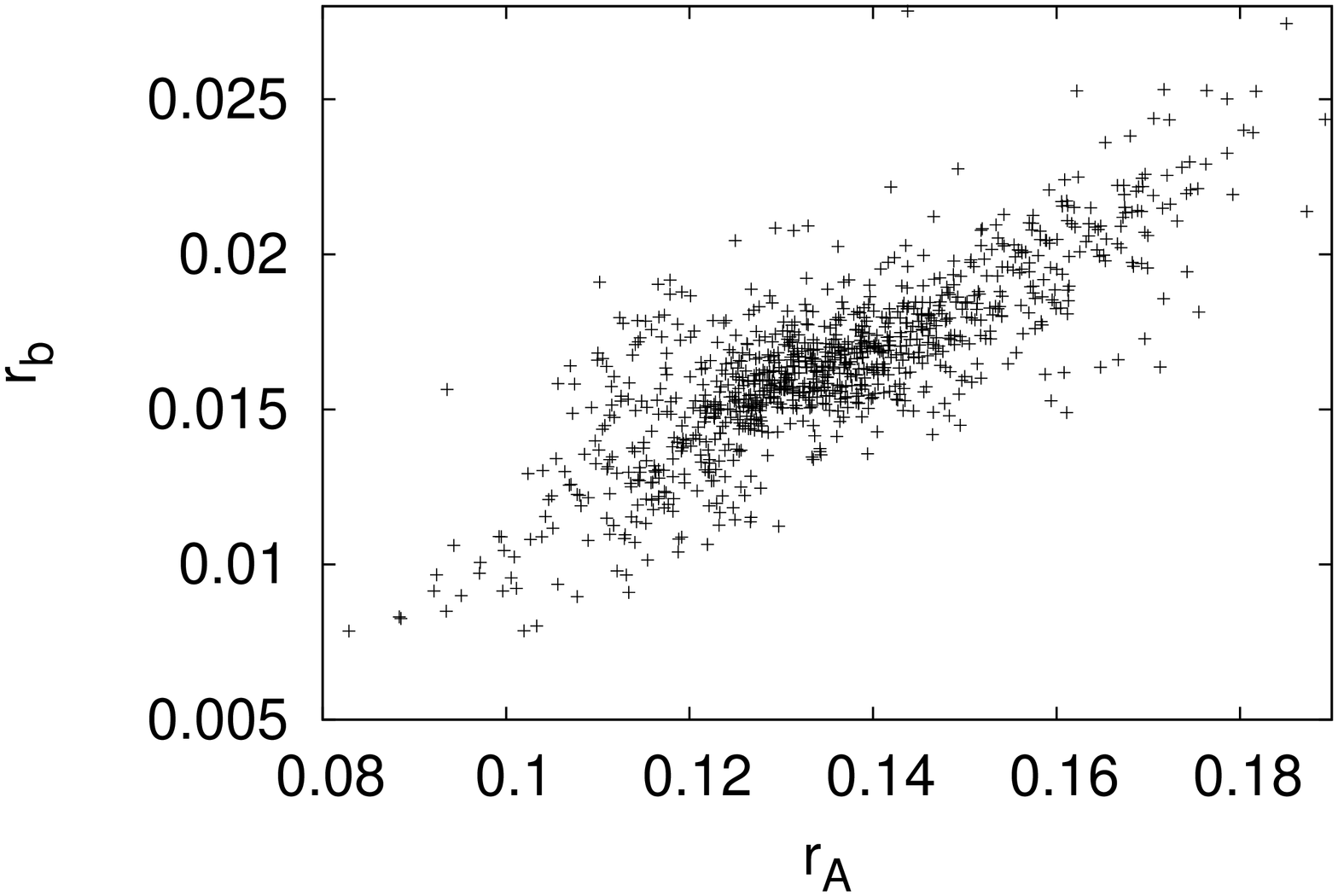}
  \includegraphics[width=0.16\textwidth,angle=270]{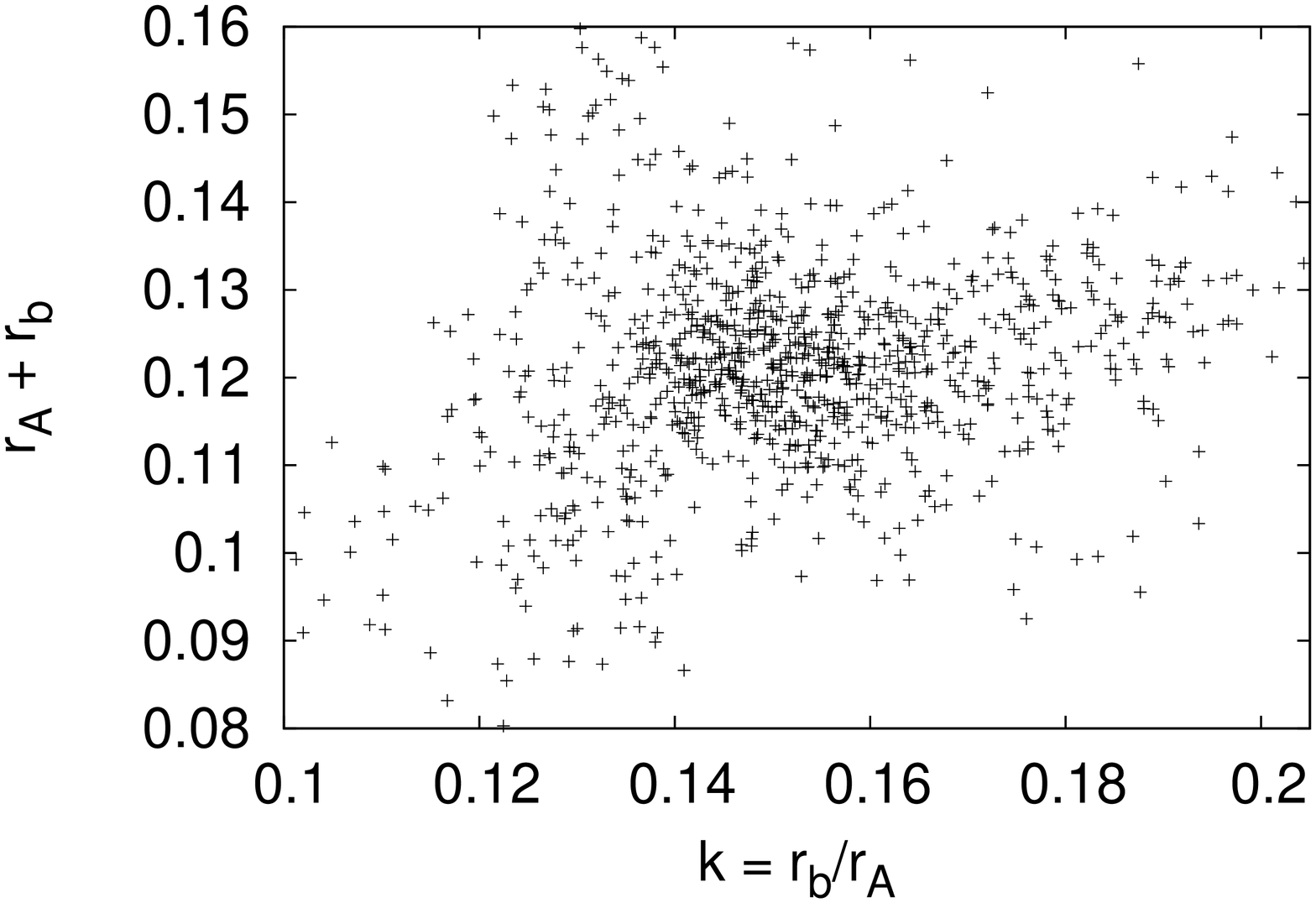} 
  \includegraphics[width=0.16\textwidth,angle=270]{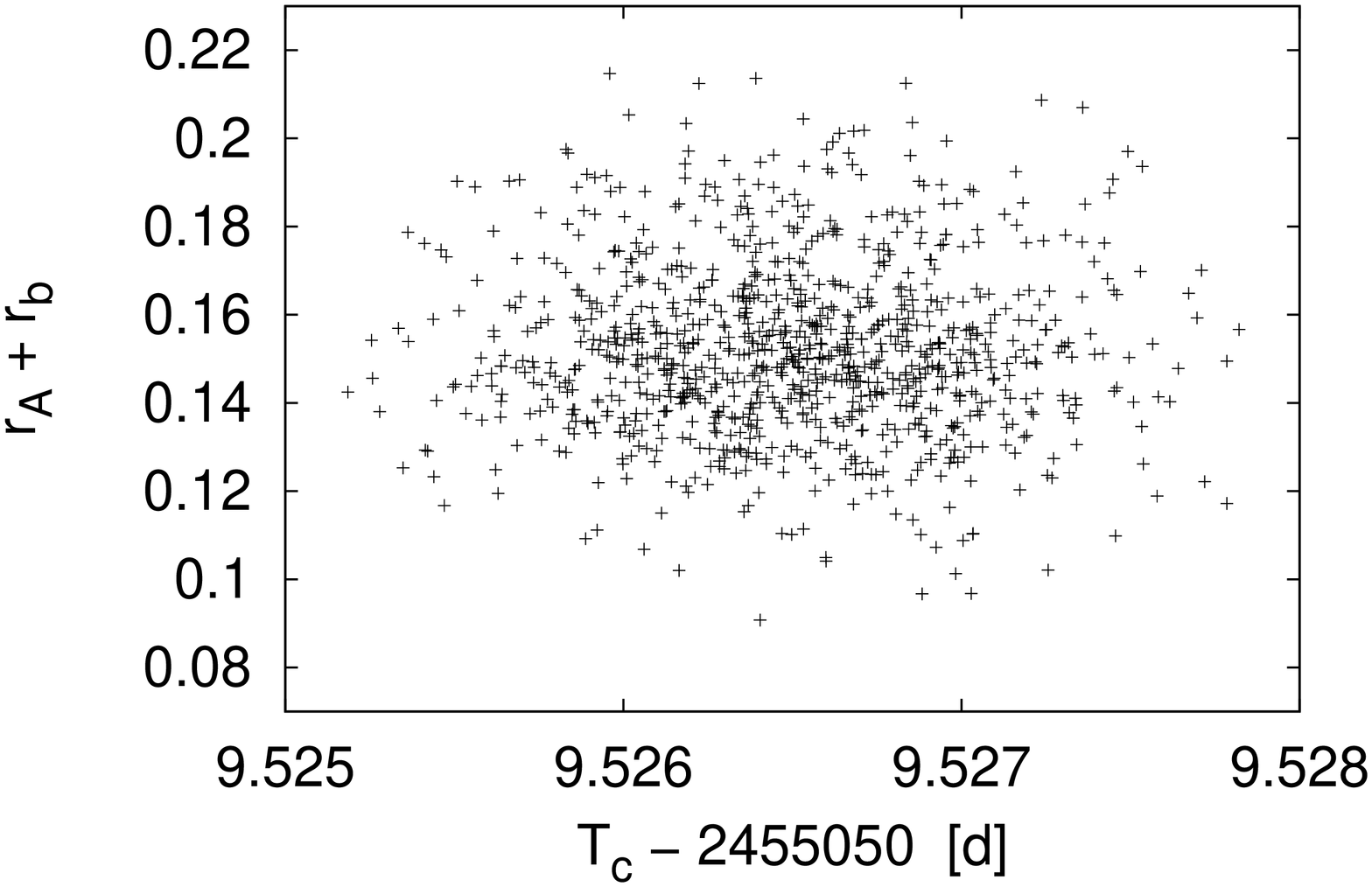} 
  \includegraphics[width=0.16\textwidth,angle=270]{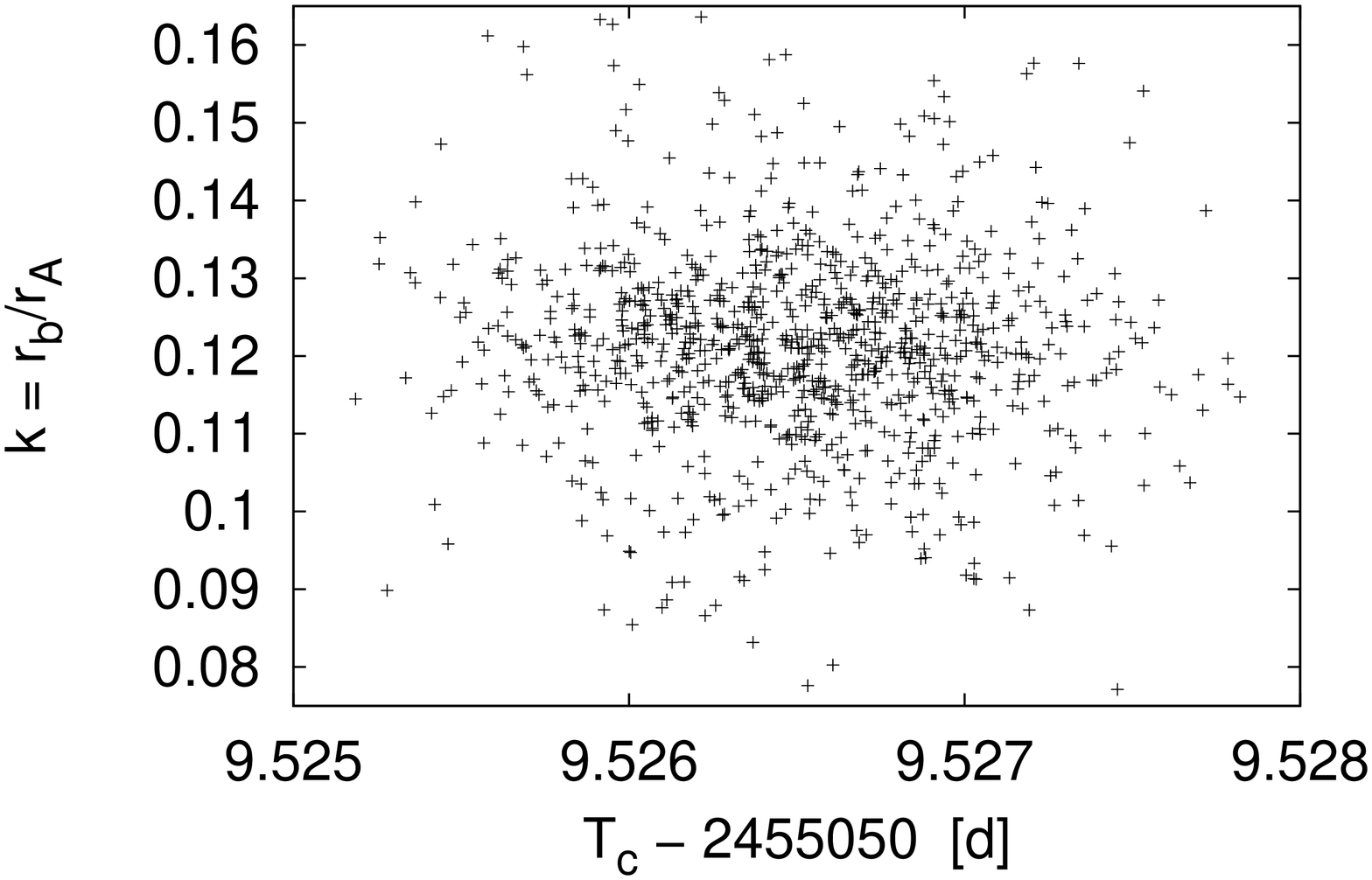} 
  \includegraphics[width=0.16\textwidth,angle=270]{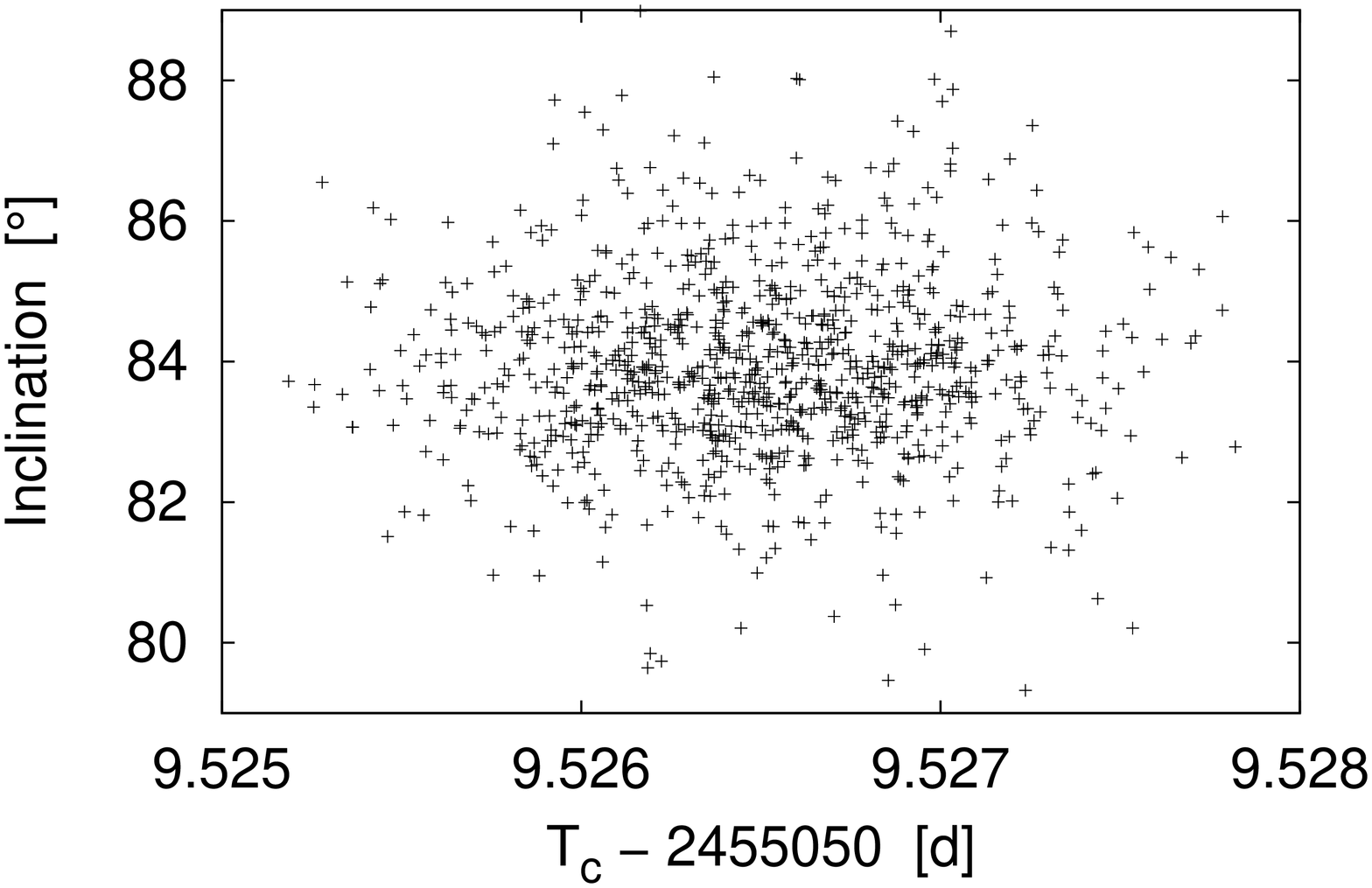}
  \includegraphics[width=0.16\textwidth,angle=270]{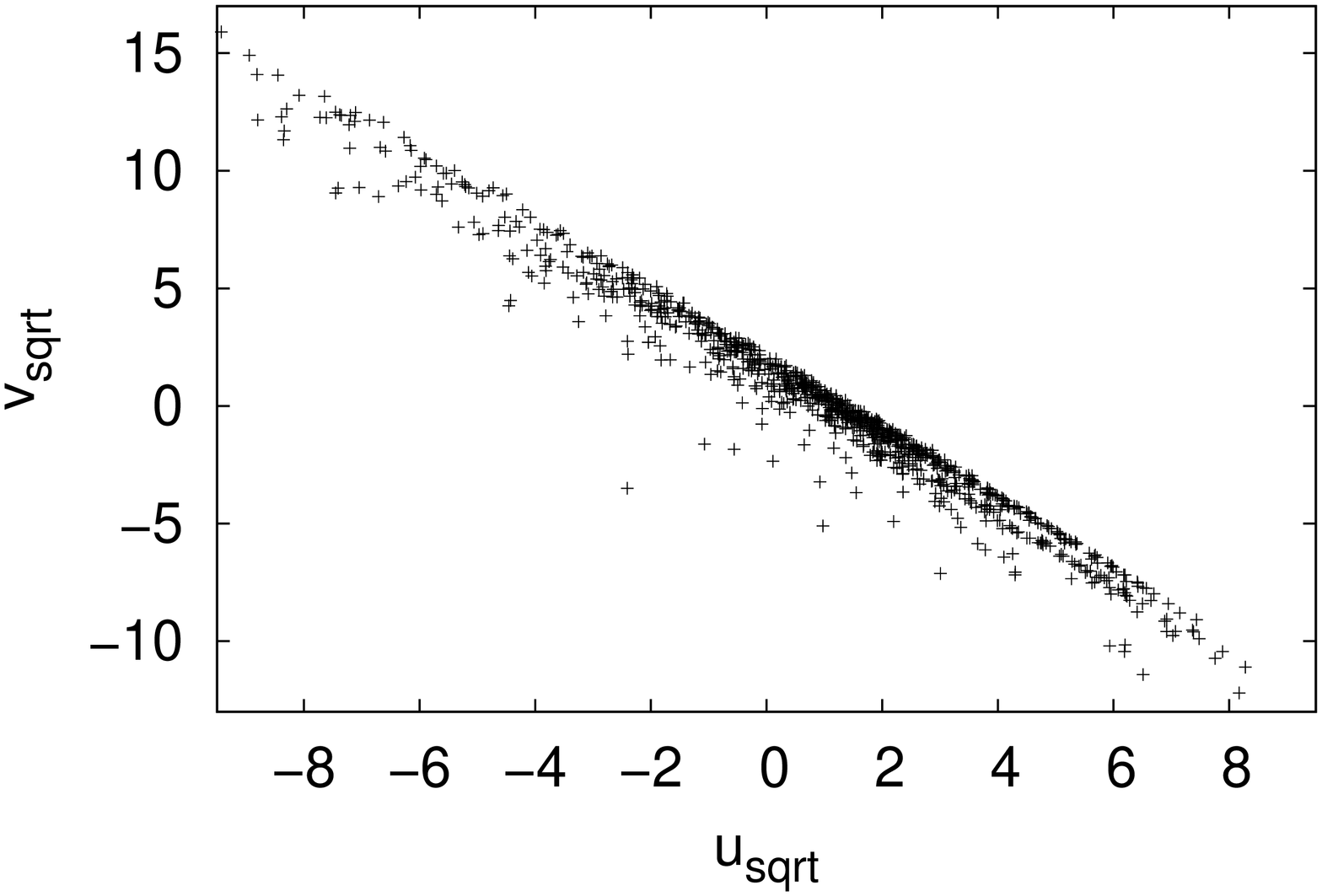}
  \includegraphics[width=0.16\textwidth,angle=270]{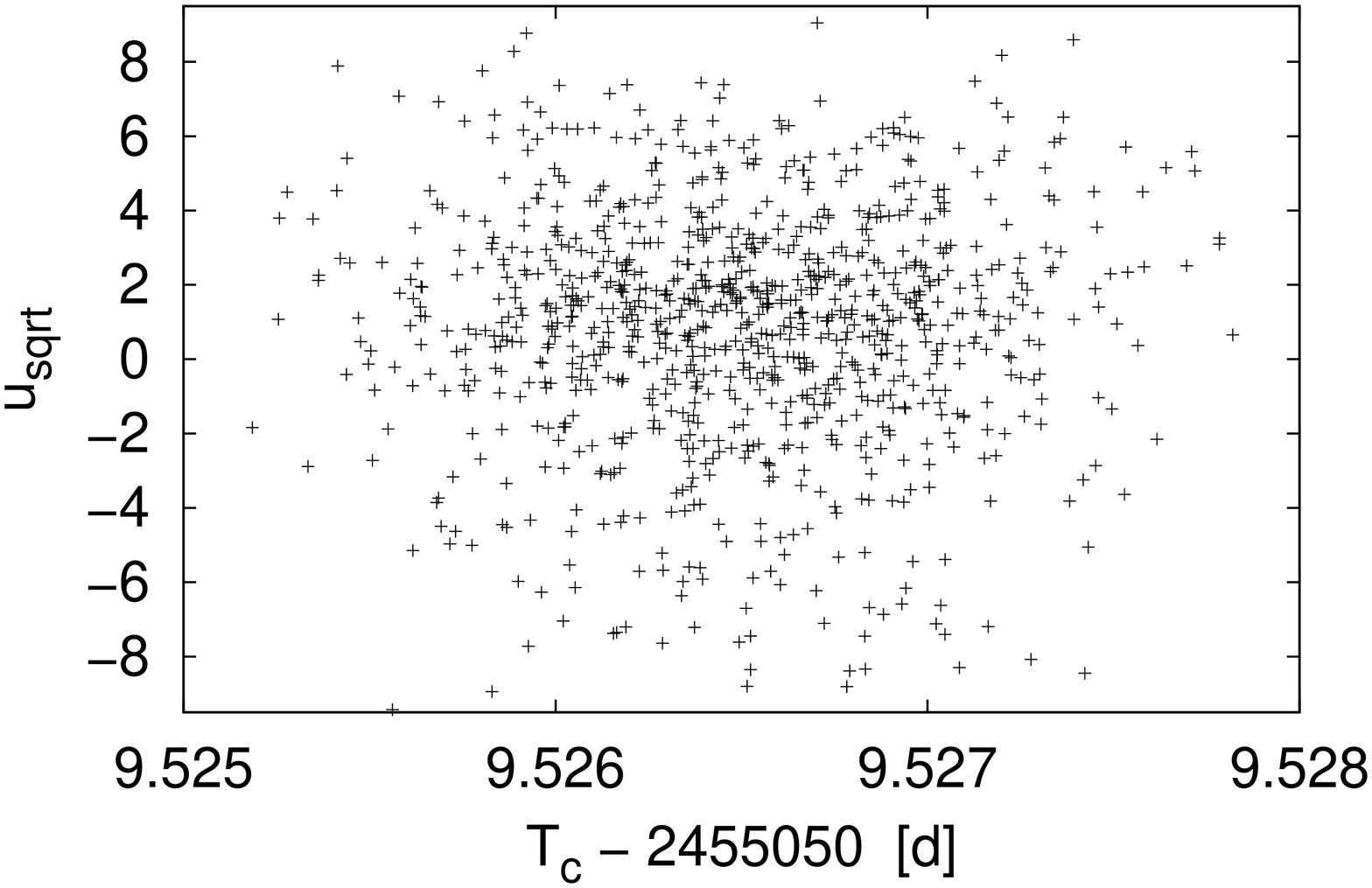}
  \includegraphics[width=0.16\textwidth,angle=270]{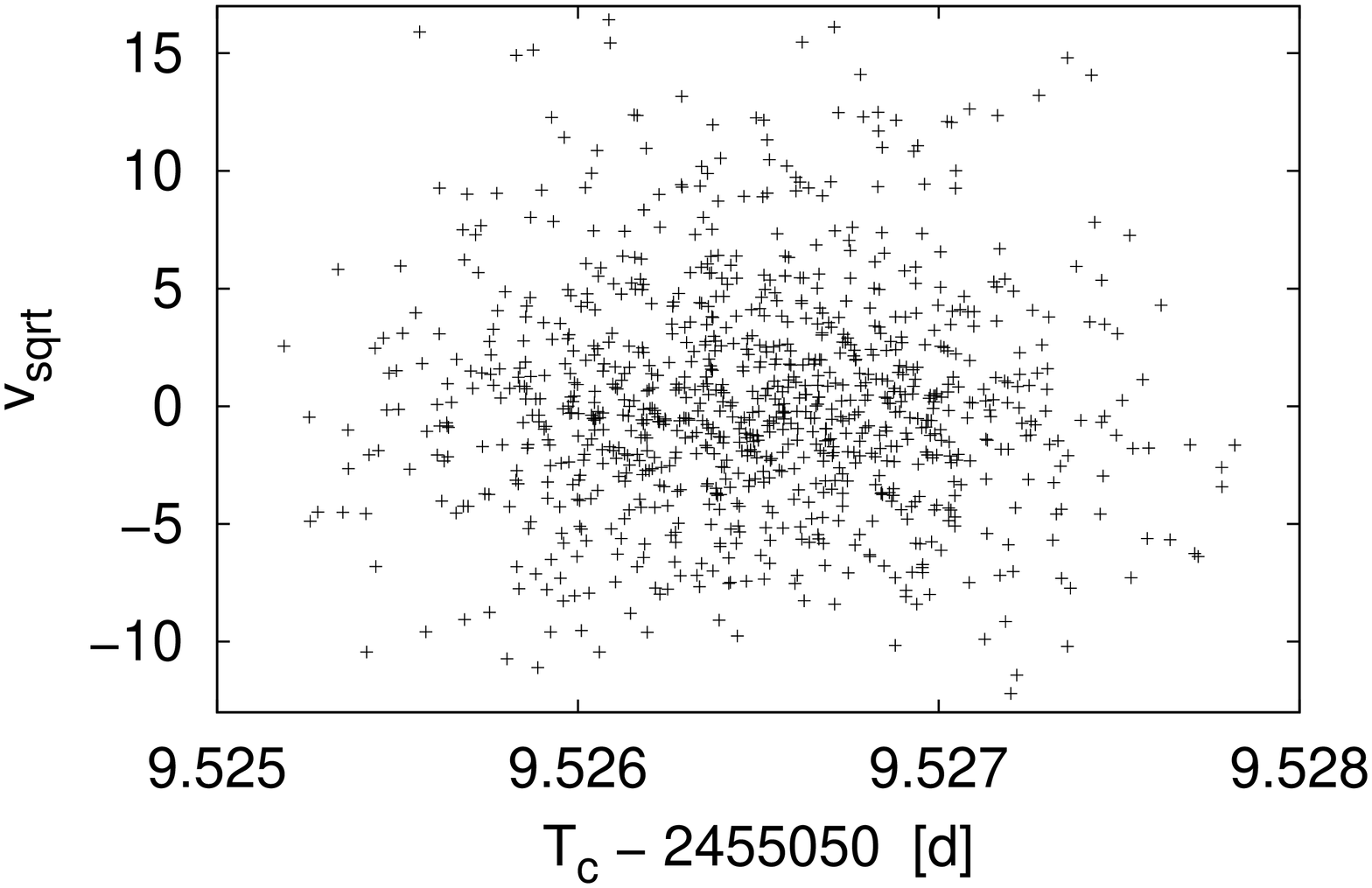}
  \includegraphics[width=0.16\textwidth,angle=270]{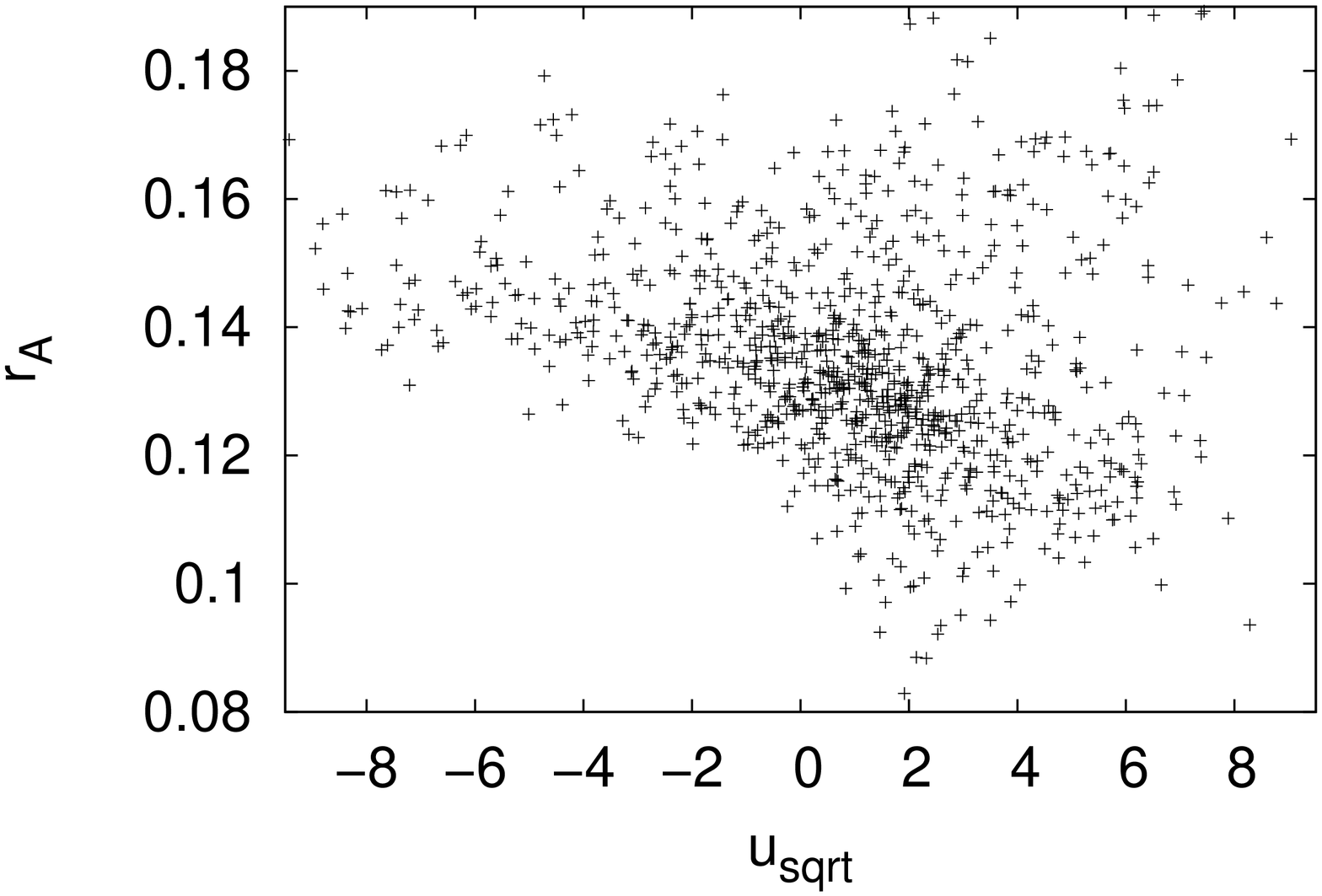}
  \includegraphics[width=0.16\textwidth,angle=270]{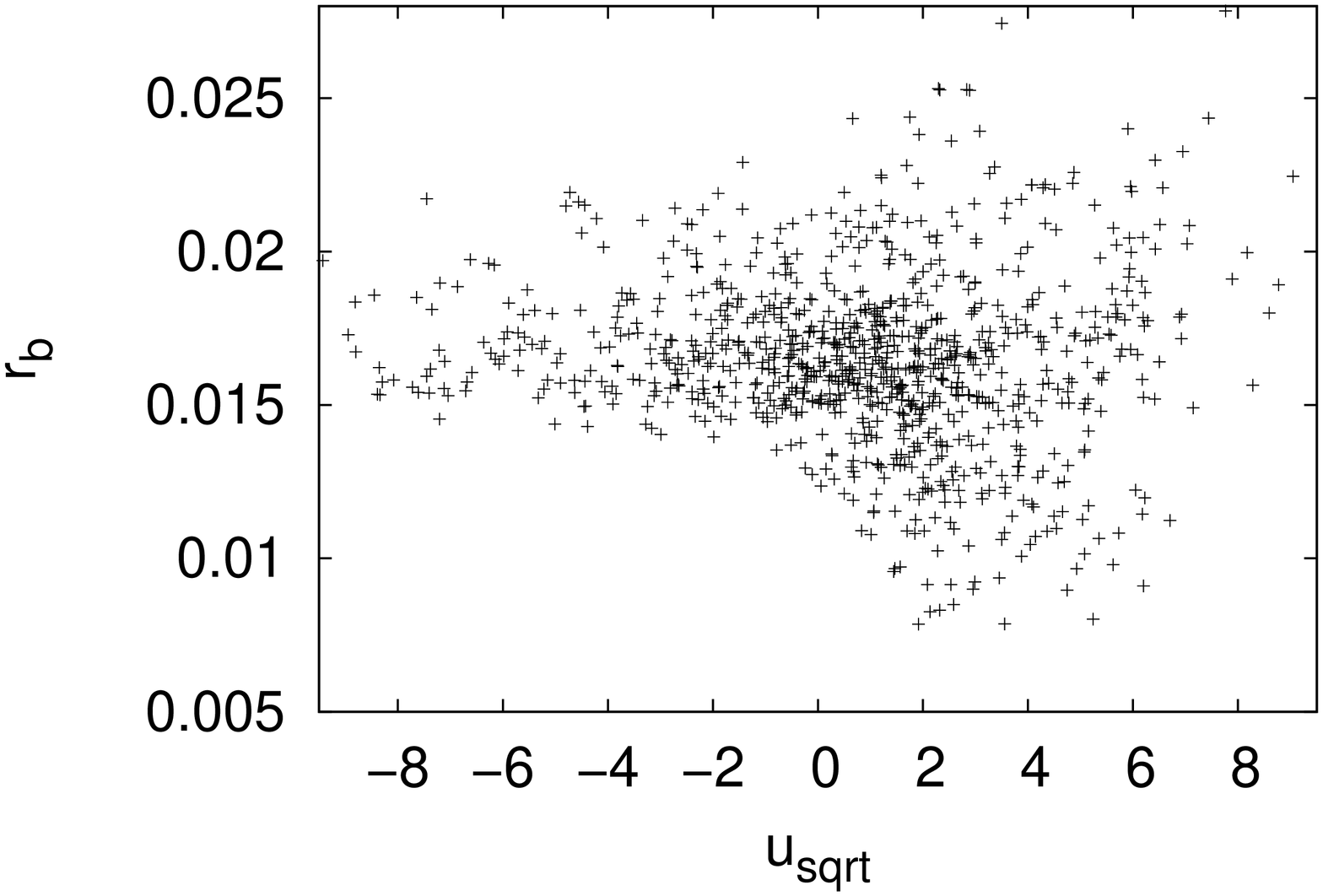}
  \includegraphics[width=0.16\textwidth,angle=270]{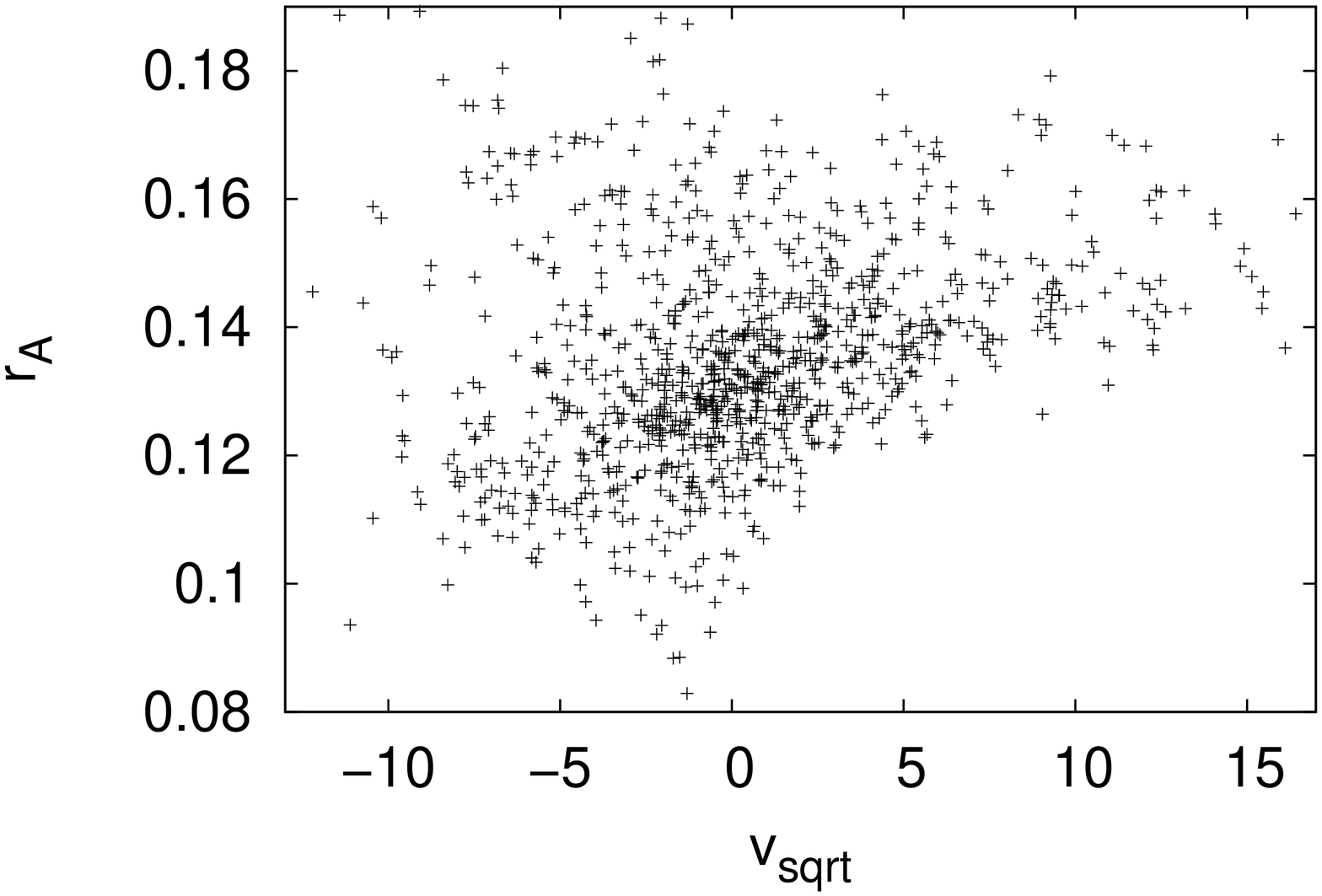}
  \includegraphics[width=0.16\textwidth,angle=270]{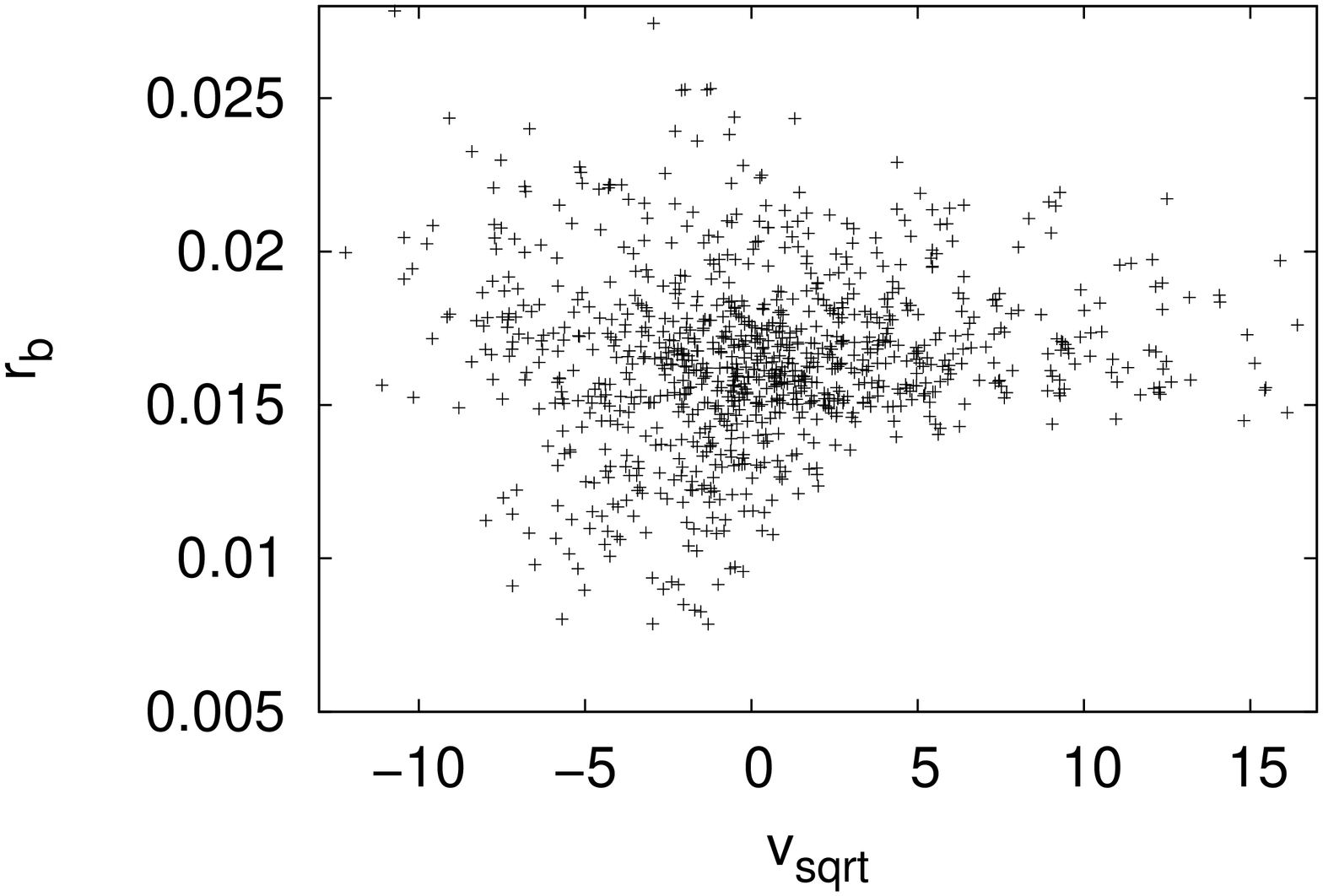}
  \caption{Scatter plots showing results from the light curve analysis of the light curve from 2009 August 15 using a Monte Carlo simulation. $r_{\rm{A}}$ and $r_{\rm{b}}$ are correlated while $k$ and $r_{\rm{A}}+r_{\rm{b}}$ show a significantly weaker correlation. Also the two LD coefficients are strongly correlated. The time of the mid-transit is not correlated to any parameter. In addition $r_{\rm{A}}$ and $r_{\rm{b}}$ show only a weak correlation to the LD coefficients. The square-root LD law was used since it is expected to show the strongest correlations \citep{2008MNRAS.386.1644S}.}
  \label{correlations}
\end{figure}
\begin{figure}
\begin{minipage}[]{0.45\textwidth}
  \centering
  \includegraphics[width=0.8\textwidth]{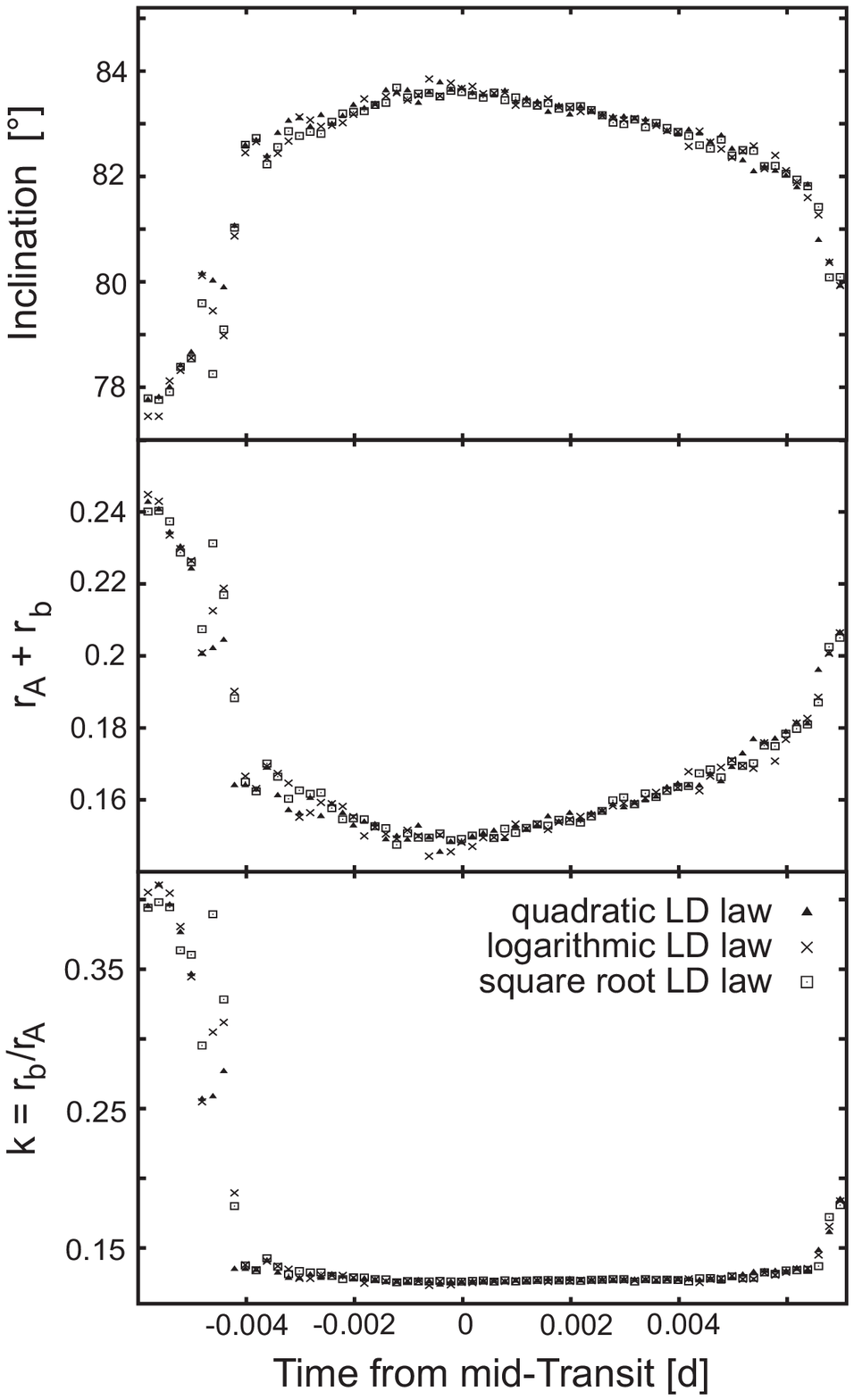}
  \caption{Inclination $i$, sum of the fractional radii ($r_{\rm{A}}+r_{\rm{b}}$) and radius ratio $k=\frac{r_{\rm{b}}}{r_{\rm{A}}}$ over time of mid-transit $T_{\mathrm{c}}$ for the light curve from 2009 August 15 for three different LD laws. $T_{\mathrm{c}}$ were kept fixed on certain values while the other parameters could vary freely. For a better distinguishability we omitted the error bars. The results for the different LD laws are very similar.}
  \label{TrES2_correlations_plots}
\end{minipage}
\begin{minipage}[]{0.45\textwidth}
  \centering
  \includegraphics[height=0.3\textheight, angle=270]{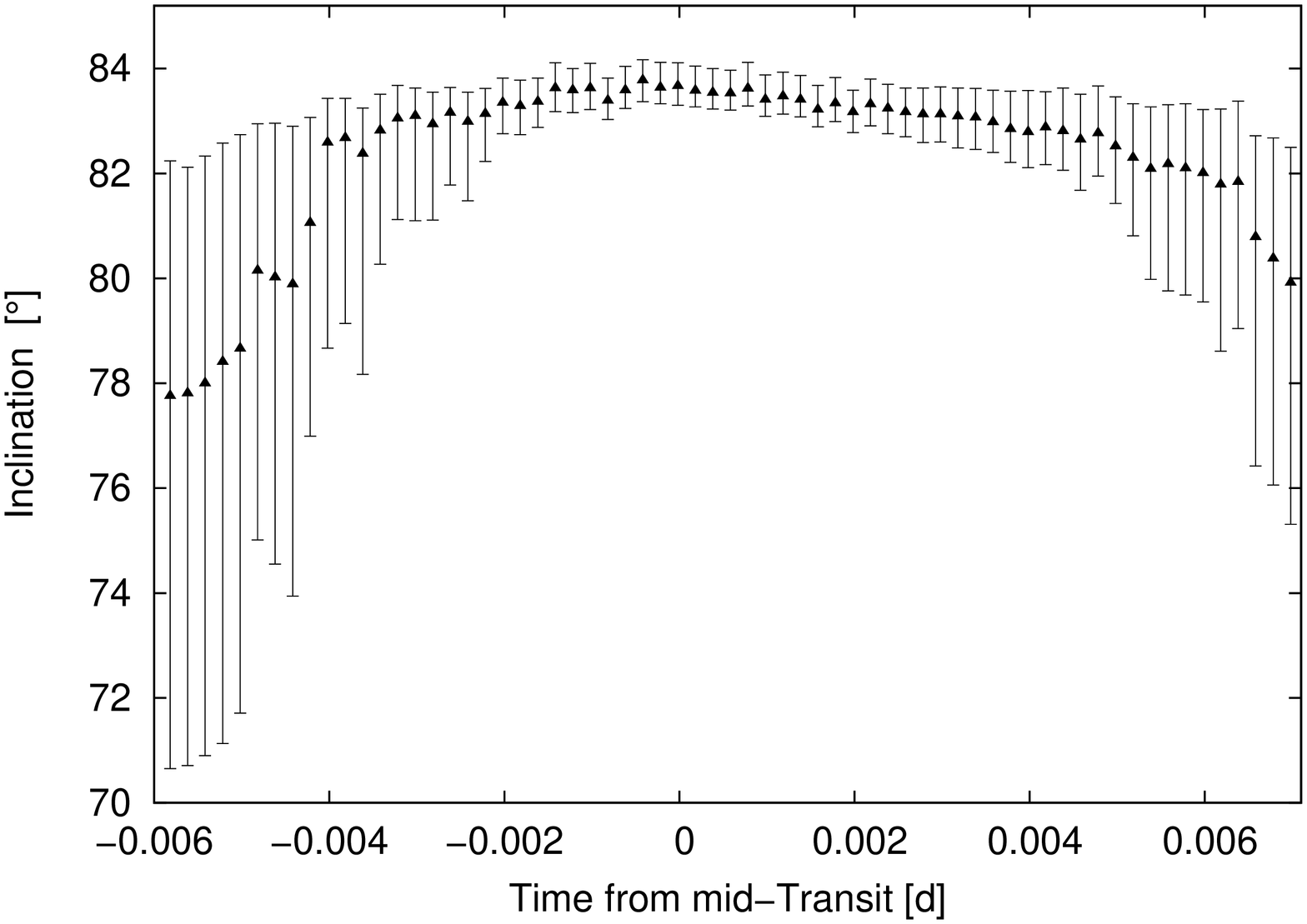}
  \caption{The same as Fig.~\ref{TrES2_correlations_plots} upper plot but only for the quadratic LD as example to show the evolution of the error bars.}
  \label{correlation_test_Inkl_quad}
\end{minipage}
\end{figure}
For the calculation of a synthetic light curve we used the \begin{scriptsize}JKTEBOP\end{scriptsize} code (version 25) \citep{2004MNRAS.349..547S,2004MNRAS.351.1277S} which is based on the EBOP program \citep{1981psbs.conf..111E,1981AJ.....86..102P} but was heavily modified (optimization algorithm, improved treatment of LD, and extensive error analysis techniques) by Southworth. In this code both components of a transit system are modelled as biaxial ellipsoids. The overall brightness as a function of the orbital period is calculated by numerical integration of concentric annuli over the surface of each body. The advantage of \begin{scriptsize}JKTEBOP\end{scriptsize} is that a small and spherical planet approximation which is usually used \citep[e.g.][]{2002ApJ...580L.171M} can be avoided. A model is fitted to the data by the Levenberg--Marquardt least-squares procedure.\\ The main parameters are the orbital period $P$, the mid-transit time  $T_{\mathrm{c}}$, the orbital inclination $i$ and the radii of star and planet expressed in relation to the semi-major axis $a$:
\begin {eqnarray}
\hspace{2cm} r_{\mathrm{A}}={R_{\mathrm{A}}\over {a}}, \hspace{2cm} r_{\mathrm{b}}={R_{\mathrm{b}} \over {a}}
\end {eqnarray}
where $R_{\mathrm{A}}$, $R_{\mathrm{b}}$ correspond to the absolute radii of star and planet, respectively.\\ \citet{2008MNRAS.386.1644S} showed that there is a very strong correlation between $r_{\rm{A}}$ and $r_{\rm{b}}$. Therefore, instead of fitting  $r_{\rm{A}}$ and $r_{\rm{b}}$ directly, \begin{scriptsize}JKTEBOP\end{scriptsize} uses parameter combinations such as the sum of the fractional radii ($r_{\rm{A}}+r_{\rm{b}}$) and the radius ratio $k=\frac{r_{\rm{b}}}{r_{\rm{A}}}$, since their correlations are significantly weaker (see Fig.~\ref{correlations} upper two plots).\\ Each light curve was modelled individually. The initial values of the parameters used for the light curve analysis are given in Table~\ref{Fitparameter_TrES2} and were calculated from the literature values in Table~\ref{Werte_TrES2}. The light curves were detrended by fitting a second-order polynomial with \begin{scriptsize}JKTEBOP\end{scriptsize} simultaneously to the modelling.\\ 
\begin{table}
\centering
\caption{Initial values for the light curve analysis of TrES-2}
\label{Fitparameter_TrES2}
\begin{tabular}{lc}
\hline \hline
Parameter & Value \\ \hline
$r_{\mathrm{A}}$ & 0.1282\,$\pm$\,0.0035 \\
$r_{\mathrm{b}}$  & 0.01658\,$\pm$\,0.00043 \\
$k=\frac{r_{\mathrm{b}}}{r_{\mathrm{A}}}$ & 0.129329 \\
$r_{\mathrm{A}}+r_{\mathrm{b}}$ & 0.14478 \\
$i$  [$^{\circ}$] & 83.62\,$\pm$\,0.14\\
$\frac{M_{\mathrm{b}}}{M_{\mathrm{A}}}$ & 0.00123 \\
\hline \hline
\end{tabular}
\end{table}
Since the quality of the observed light curves is very different, some parameters had to be fixed to the initial values in cases of light curves with very low S/N, or partial transits, otherwise unphysical results were generated. The parameters that were fixed for each light curve are given in Table~\ref{LCA_ground}. To detect possible transit time variations, the mid-transit time $T_{\mathrm{c}}$ was always a free parameter, whose initial value was calculated using the ephemeris of \citet{2009AN....330..459R}.\\ To test different LD laws we modelled every light curve with a quadratic, logarithmic, and square-root law. The theoretical LD coefficients were bilinearly interpolated from the tables by \citet{2000A&A...363.1081C} and \citet{2010A&A...510A..21S} using the stellar parameters in Table~\ref{Werte_TrES2} and are given in Table~\ref{LD_TrES2}. Since the LD coefficients can only be determined on transit observations with a very high S/N, we fixed the values to the ones from literature (Table~\ref{LD_TrES2}). To include the contribution of the LD coefficients to the error budget they were perturbed by $\pm0.1$ around the theoretical values assuming a flat distribution. As \citet{2008MNRAS.386.1644S} showed, strong correlations can be seen between the two LD coefficients for all three laws which is expected in respect to the quality of the ground-based light curves. One example is given in Fig.~\ref{correlations} for the square-root law. Parameters $r_{\rm{A}}$ and $r_{\rm{b}}$ show no significant correlation to the LD coefficients (see Fig.~\ref{correlations} lower two panels) meaning that their values have very little effect on the light curve modelling. \\ Selection criterion for the LD law was the standard deviation $\sigma$ of the residuals of the individual fits. The LD law that produced the lowest $\sigma$ was chosen for the final parameter determination (the chosen LD law for each light curve is given in Table~\ref{LCA_ground}). One example is given in Table~\ref{result_09_08_15}. To test our method, first we fixed both LD coefficients to the theoretical values, then only the nonlinear LD coefficient were kept fixed and finally both coefficients were allowed to vary. A comparison of the results of the different approaches in Table~\ref{result_09_08_15} confirms the conclusions of \citet{2008MNRAS.386.1644S}. If both LD coefficients were kept fixed, the smallest errors are obtained. Fixing only one of the coefficients can produce unphysical results even with a light curve with relatively high quality. If both coefficients are allowed to vary, unrealistic results with very large errors are obtained. Within the error bars the mid-transit time is consistent in each of the nine scenarios in Table~\ref{result_09_08_15}. For the light curve from 2009 August 15 the quadratic LD law produces the lowest $\sigma$. Therefore the first column of Table~\ref{result_09_08_15} represents the final results for this transit. In general, the achieved precision is not sufficient to distinguish between the LD laws, as $\sigma$ only varies in the second or third decimal place. Nevertheless, we considered the synthetic light curve which produces the residuals with the smallest $\sigma$ to be the best fit to the observed data. The choice of the LD law does not influence the mid-transit times. The resulting transit times $T_{\mathrm{c}}$ are consistent within the error bars as shown in Table~\ref{result_09_08_15}. $T_{\mathrm{c}}$ is not correlated with $r_{\mathrm{A}}+r_{\mathrm{b}}$, $k$, $i$, or the LD coefficients as shown in Fig.~\ref{correlations}. To check how the system parameters evolve with the time of mid-transit under different LD laws we did multiple light curve modelling with fixed value of $T_{\mathrm{c}}$ while the other parameters could vary freely. We used 70 values $\pm9$\,min around $T_{\mathrm{c}}$. The result is shown in Fig.~\ref{TrES2_correlations_plots}. The evolution of the error bars is given in Fig.~\ref{correlation_test_Inkl_quad}. The best fitting values for the parameters are always the extreme points (maximum for $i$ and minimum for $r_{\mathrm{A}}+r_{\mathrm{b}}$, $k$) in the diagram at the time $T_{\mathrm{c}}$. These values also have the smallest uncertainties. Although the scatter is smaller for the quadratic LD law (example for $i$: quadratic LD - $\sigma=1.57^{\circ}$, logarithmic LD - $\sigma=1.64^{\circ}$, square-root LD - $\sigma=1.65^{\circ}$), within the error bars all three LD laws are indistinguishable. \\ To determine parameter errors we applied two different methods for each light curve. To calculate random errors of light curves with Gaussian noise we ran 1000 Monte Carlo simulations. However, light curves obtained from ground-based telescopes are affected by systematic effects which produce noise on time-scales similar to the duration of the transit \citep{2006MNRAS.373..231P}. This so-called 'red noise' leads to a correlation between adjacent data points. To account for 'red noise' we used a residual permutations algorithm \citep['prayer bead',][]{2002ApJ...564..495J}  which is implemented in \begin{scriptsize}JKTEBOP\end{scriptsize} \citep{2005MNRAS.363..529S}. In the case of light curves with systematic effects the Monte Carlo errors were found to be 2--3 times smaller than the values returned by the prayer bead method. Thus, for light curves with ''red noise'' we used the prayer bead errors as the final ones. In cases where the errors of both methods are comparable we used the Monte Carlo errors. This treatment leads to an estimation of robust and realistic error bars. \\ Our final light curves are are given at the end of this paper where the solid line gives the best fitting theoretical light curve. The results of the fitting of the individual light curves are given in Table~\ref{LCA_ground}. \\ In our first observations of a transit of TrES-2 \citep{2009AN....330..459R} we could detect a brightness drop ('dip') $\sim$\,1-2\,h after the end of the transit which was also seen in one light curve of \citet{2006ApJ...651L..61O}. We tried to explain its existence with five different theories: a nearby variable star or a blended eclipsing binary, an additional planet in the system, a transit over a background star, a transit over a wide companion of the TrES-2 host star after the actual transit is finished or a photometric false alarm. In none of our observed light curve we could find evidence for a reoccurrence of the 'dip'.

\begin{table}
\centering
\caption{Theoretical LD coefficients for TrES-2 from the tables by \citet{2000A&A...363.1081C} and \citet{2010A&A...510A..21S} for $V$, $R$, $I$, and Kepler wavelength, respectively.}
\label{LD_TrES2}
\begin{small}\begin{tabular}{ccccccc}
\hline \hline
Filter & \multicolumn{2}{c}{Quadratic} & \multicolumn{2}{c}{Logarithmic} & \multicolumn{2}{c}{Square-root}  \\ 
& $u_{\mathrm{q}}$ & $v_{\mathrm{q}}$ & $u_{\mathrm{lg}}$ & $v_{\mathrm{lg}}$ & $u_{\mathrm{s}}$ & $v_{\mathrm{s}}$ \\ \hline
$V\mathrm{j}$ & 0.3955 & 0.3203 & 0.7448 & 0.2639 & 0.2646 & 0.5324 \\
$R\mathrm{c}$ & 0.3051 & 0.3388 & 0.6751 & 0.2808 & 0.1621 & 0.5692 \\
$I\mathrm{c}$ & 0.2293 & 0.3326 & 0.5932 & 0.2777 & 0.0832 & 0.5664 \\ 
\textit{Kepler} & 0.3622 & 0.2795 & & & & \\\hline \hline
\end{tabular}\end{small}
\end{table}

\begin{table}
\caption{Parameters of light curve analysis for the transit on 2009 Aug. 15 for three different LD laws. $T_{\mathrm{c}}$ is based on UTC and is given as 2455059+.}
\label{result_09_08_15}
\renewcommand{\arraystretch}{1.3} 
\begin{scriptsize}\begin{tabular}{cccc}
\hline \hline
Parameter & Quadratic law & Logarithmic law & Square-root law \\ \hline 
\multicolumn{4}{l}{All LD coefficients fixed*} \\ \hline 
$r_{\mathrm{A}}+r_{\mathrm{b}}$ &  $0.1494^{+0.0079}_{-0.0077}$ & $0.1507^{+0.0084}_{-0.0077}$ & $0.1506^{+0.0073}_{-0.0073}$ \\
$k$ & $0.1241^{+0.0041}_{-0.0033}$ & $0.1248^{+0.0038}_{-0.0035}$ & $0.1252^{+0.0037}_{-0.0032}$ \\
$i$  [$^{\circ}$] & $83.59^{+0.50}_{-0.48}$ & $83.50^{+0.47}_{-0.47}$ & $83.49^{+0.44}_{-0.42}$ \\ 
$T_{\mathrm{c}}$  [d] & $0.52692^{+0.00049}_{-0.00047}$ & $0.52690^{+0.00046}_{-0.00048}$ & $0.52678^{+0.00049}_{-0.00048}$ \\
$u$ & 0.3955 (fixed*) & 0.7448 (fixed*) & 0.2646 (fixed*) \\
$v$ & 0.3202 (fixed*) & 0.2639 (fixed*) & 0.5324 (fixed*) \\
$r_{\mathrm{A}}$ & $0.1330^{+0.0066}_{-0.0067}$ & $0.1338^{+0.0070}_{-0.0063}$ & $0.1339^{+0.0060}_{-0.0063}$ \\
$r_{\mathrm{b}}$ & $0.0165^{+0.0013}_{-0.0012}$ & $0.0167^{+0.0013}_{-0.0012}$ & $0.0167^{+0.0012}_{-0.0010}$ \\
$\sigma$  [mmag] & 2.7340 & 2.7342 & 2.7365 \\\hline 
\multicolumn{4}{l}{Only nonlinear LD coefficient fixed*} \\ \hline 
$r_{\mathrm{A}}+r_{\mathrm{b}}$ &  $0.1553^{+0.0190}_{-0.0125}$ & $0.1573^{+0.0209}_{-0.0124}$ & $0.1567^{+0.0205}_{-0.0115}$ \\
$k$ & $0.1229^{+0.0065}_{-0.0082}$ & $0.1242^{+0.0071}_{-0.0079}$ & $0.1250^{+0.0069}_{-0.0079}$ \\
$i$  [$^{\circ}$] & $83.56^{+0.89}_{-0.80}$ & $83.45^{+0.79}_{-0.93}$ & $83.49^{+0.70}_{-0.90}$ \\ 
$T_{\mathrm{c}}$  [d] & $0.52683^{+0.00047}_{-0.00054}$ & $0.52665^{+0.00052}_{-0.00056}$ & $0.52663^{+0.00050}_{-0.00054}$ \\
$u$ &  $0.87^{+0.29}_{-0.65}$ &  $1.22^{+0.29}_{-0.78}$ &  $0.74^{+0.27}_{-0.67}$ \\
$v$ & 0.3202 (fixed*) & 0.2639 (fixed*) & 0.5324 (fixed*) \\
$r_{\mathrm{A}}$ & $0.1382^{+0.0163}_{-0.0105}$ & $0.1402^{+0.0176}_{-0.0108}$ & $0.1393^{+0.0176}_{-0.0101}$ \\
$r_{\mathrm{b}}$ & $0.0170^{+0.0025}_{-0.0020}$ & $0.0173^{+0.0029}_{-0.0019}$ & $0.0173^{+0.0030}_{-0.0017}$ \\
$\sigma$  [mmag] & 2.7365 & 2.7403 & 2.7407 \\\hline
\multicolumn{4}{l}{Fitting for both LD coefficients} \\ \hline 
$r_{\mathrm{A}}+r_{\mathrm{b}}$ &  $0.1558^{+0.0193}_{-0.0134}$ & $0.1544^{+0.0180}_{-0.0199}$ & $0.1579^{+0.0216}_{-0.0173}$ \\
$k$ & $0.1135^{+0.0148}_{-0.0094}$ & $0.1163^{+0.0120}_{-0.0109}$ & $0.1229^{+0.0122}_{-0.0117}$ \\
$i$  [$^{\circ}$] & $83.32^{+1.15}_{-0.97}$ & $83.53^{+1.42}_{-0.93}$ & $83.53^{+1.28}_{-1.05}$ \\ 
$T_{\mathrm{c}}$  [d] & $0.52684^{+0.00049}_{-0.00053}$ & $0.52679^{+0.00051}_{-0.00056}$ & $0.52654^{+0.00053}_{-0.00056}$ \\
$u$ &  $-0.60^{+2.40}_{-6.27}$ &  $1.38^{+1.53}_{-1.39}$ &  $1.37^{+3.10}_{-3.72}$ \\
$v$ &  $1.75^{+7.11}_{-2.87}$ &  $0.79^{+4.02}_{-2.52}$ &  $-0.49^{+4.97}_{-4.77}$ \\
$r_{\mathrm{A}}$ & $0.1398^{+0.0162}_{-0.0129}$ & $0.1384^{+0.0167}_{-0.0191}$ & $0.1406^{+0.0182}_{-0.0147}$ \\
$r_{\mathrm{b}}$ & $0.0160^{+0.0025}_{-0.0017}$ & $0.0160^{+0.0023}_{-0.0019}$ & $0.0172^{+0.0031}_{-0.0023}$ \\
$\sigma$  [mmag] & 2.7405 & 2.7414 & 2.7526 \\ \hline \hline
\end{tabular}\end{scriptsize}
$^{\ast}$Permuted by $\pm$\,0.1 on a flat distribution
\end{table}

\begin{table*}
\caption{Results of the light curve analysis for the ground-based transits. The values without error bars are the initial values from Table~\ref{Fitparameter_TrES2} that were fixed in the analysis but permuted by $\pm$\,0.1 on a flat distribution to include their contribution to the error budget. The LD coefficients were always kept fixed on the theoretical values in Table~\ref{LD_TrES2}. The detrending coefficients that were used are given as Poly-coeff. 1 and Poly-coeff. 2. If multiple transits per epoch are available the indices mark the following instruments (description in Table~\ref{CCD_Kameras}): (1) CTK, (2) RTK, (3) STK, (4) CTK-II, (5) MONICA, (6) ST6, (7) G2-1600, (8) Trebur.}
\label{LCA_ground}
\begin{small}\begin{tabular}{lccccr@{\,$\pm$\,}lr@{\,$\pm$\,}l}
\hline \hline
Epoch & $r_{\mathrm{A}}+r_{\mathrm{b}}$ & $k$ & $i$ [$^{\circ}$] & LD law$^{(a)}$ & \multicolumn{2}{c}{Poly-coeff. 1 (x)} & \multicolumn{2}{c}{Poly-coeff. 2 (x$^{2}$)} \\ \hline 
87	& $0.1432^{+0.0063}_{-0.0062}$ & $0.1290$ & $83.62$ & quad &                                      2.05	&	2.61	&	-0.42	&	0.50\\
108	& $0.1454^{+0.0106}_{-0.0074}$ & $0.1290$ & $83.62$ & log &                                    -2.39	&	3.16	&	0.26	&	0.35\\
138	& $0.1450$ & $0.1390^{+0.0370}_{-0.0200}$ & $83.62$ & quad &                                    15.33	&	4.62	&	-0.89	&	0.27\\
142	& $0.1447^{+0.0451}_{-0.0369}$ & $0.1162^{+0.0263}_{-0.0194}$ & $83.85^{+2.98}_{-2.08}$ & sqrt & 0.00	&	3.27	&	0.00	&	0.19\\
163$^{(1)}$	& $0.1450$ & $0.1264^{+0.0141}_{-0.0126}$ & $83.74^{+0.46}_{-0.38}$ &  sqrt &                   0.47	&	0.35	&	-0.63	&	0.41\\
163$^{(6)}$	& $0.1450$ & $0.1290$ & $83.62$ & quad &                                                        0.91	&	0.10	&	-1.05	&	0.12\\
163$^{(5)}$	& $0.1294^{+0.0182}_{-0.0161}$ & $0.1227^{+0.0067}_{-0.0073}$ & $84.55^{+1.01}_{-1.10}$ & sqrt & 0.14	&	0.08	&	-0.16	&	0.11\\
165	& $0.1450$ & $0.1439^{+0.0212}_{-0.0157}$ & $83.62$ &   sqrt &                                  2.42	&	2.00	&	-0.22	&	0.18\\
174	& $0.1450^{+0.0567}_{-0.0358}$ & $0.1414^{+0.2281}_{-0.0176}$ & $83.69^{+2.25}_{-3.81}$ &  log & -6.36	&	6.30	&	0.42	&	0.42\\
278	& $0.1338^{+0.0112}_{-0.0109}$ & $0.1290$ & $83.62$ & quad &                                     9.07	&	2.87	&	-1.03	&	0.32\\
316	& $0.1450$ & $0.1293^{+0.0161}_{-0.0111}$ & $83.62$ & quad &                                     0.00	&	7.30	&	0.00	&	0.44\\
318	& $0.1450$ & $0.1214^{+0.3651}_{-0.0255}$ & $83.15^{+1.15}_{-0.83}$ & quad &                     0.00	&	87.17	&	0.00	&	13.08\\
395	& $0.1432^{+0.0071}_{-0.0054}$ & $0.1392^{+0.0271}_{-0.0163}$ & $83.62$ &   sqrt &              6.44	&	2.01	&	-0.88	&	0.28\\
399	& $0.1450$ & $0.1290$ & $83.58^{+0.35}_{-0.24}$ &  quad &                                        1.67	&	0.98	&	-0.25	&	0.14\\
414	& $0.1482^{+0.0036}_{-0.0043}$ & $0.1290$ & $83.62$ &   log &                                    0.08	&	0.25	&	-0.09	&	0.26\\
416	& $0.1102^{+0.0355}_{-0.0359}$ & $0.1290$ & $85.28^{+1.17}_{-1.80}$ &  log &                     0.00	&	6.75	&	0.00	&	0.62\\
446$^{(7)}$	& $0.1450$ & $0.1290$ & $84.12^{+0.65}_{-0.60}$ &   sqrt &                                      -14.28	&	1.86	&	0.75	&	0.10\\
446$^{(2)}$	& $0.1450$ & $0.1290$ & $83.76^{+0.29}_{-0.29}$ &   sqrt &                                      0.00	&	9.28	&	0.00	&	0.49\\
446$^{(3)}$	& $0.1502^{+0.0049}_{-0.0071}$ & $0.1241^{+0.0074}_{-0.0057}$ & $83.54^{+0.46}_{-0.31}$ & quad & 0.00	&	1.97	&	0.00	&	0.10\\
448$^{(1)}$	& $0.1450$ & $0.1290$ & $83.62$ & quad &                                                         0.12	&	0.09$^{(b)}$	&	\multicolumn{2}{c}{}\\
448$^{(2)}$	& $0.1450$ & $0.1290$ & $83.62$ &  sqrt &                                                       -0.08	&	0.07$^{(b)}$	&	\multicolumn{2}{c}{}\\
448$^{(3)}$	& $0.1450$ & $0.1177^{+0.0800}_{-0.0424}$ & $83.62$ & quad &                                     0.00	&	42.82	&	0.00	&	4.83\\
452	& $0.1395^{+0.0157}_{-0.0221}$ & $0.1271^{+0.0118}_{-0.0136}$ & $84.29^{+1.17}_{-0.87}$ & sqrt & 0.00	&	12.01	&	0.00	&	1.37\\
463$^{(2)}$	& $0.1166^{+0.1424}_{-0.0257}$ & $0.1337^{+0.0611}_{-0.0200}$ & $87.29^{+2.54}_{-6.26}$ & quad & 0.00	&	12.08	&	0.02	&	3.87\\
463$^{(3)}$	& $0.1469^{+0.0175}_{-0.0208}$ & $0.1264^{+0.0101}_{-0.0085}$ & $83.86^{+1.31}_{-1.00}$ &   log &0.00	&	1.25	&	0.01	&	0.40\\
488	& $0.1469^{+0.0033}_{-0.0056}$ & $0.1290$ & $83.62$ &   log &                                    0.00	&	2.83	&	0.00	&	0.43\\
548	& $0.1450$ & $0.1290$ & $83.62$ &  sqrt &                                                       -0.71	&	1.36	&	0.18	&	0.44\\
567	& $0.1496^{+0.0067}_{-0.0057}$ & $0.1290$ & $83.62$ &  log &                                     0.01	&	15.05	&	0.00	&	0.89\\
586	& $0.1450$ & $0.1188^{+0.0166}_{-0.0119}$ & $83.62$ & quad &                                     0.01	&	5.45	&	-0.01	&	0.50\\
622$^{(4)}$	& $0.1456^{+0.0049}_{-0.0053}$ & $0.1343^{+0.0198}_{-0.0110}$ & $83.62$ & quad &                 18.29	&	5.94	&	-2.12	&	0.68\\
622$^{(3)}$	& $0.1403^{+0.0070}_{-0.0107}$ & $0.1246^{+0.0046}_{-0.0037}$ & $84.09^{+0.55}_{-0.46}$ &  log & 0.00	&	1.69	&	0.00	&	0.19\\
622$^{(8)}$	& $0.1395^{+0.0067}_{-0.0068}$ & $0.1199^{+0.0094}_{-0.0102}$ & $84.08^{+0.51}_{-0.49}$ & sqrt & 1.18	&	0.88	&	-0.13	&	0.10\\
624	& $0.1450$ & $0.1290$ & $83.62$ & sqrt &                                                        0.00	&	16.18	&	0.00	&	0.87\\
682$^{(4)}$	& $0.1419^{+0.0208}_{-0.0295}$ & $0.1432^{+0.0189}_{-0.0151}$ & $84.36^{+1.86}_{-1.25}$ &  log & -6.47	&	5.43	&	1.23	&	1.04\\
682$^{(3)}$	& $0.1426^{+0.0084}_{-0.0087}$ & $0.1305^{+0.0058}_{-0.0046}$ & $83.85^{+0.48}_{-0.51}$ & sqrt & -0.01	&	1.42	&	0.00	&	0.27\\
684$^{(4)}$	& $0.1450$ & $0.1290$ & $83.62$ &  sqrt &                                                       0.00	&	13.75	&	0.00	&	0.91\\
684$^{(3)}$	& $0.1305^{+0.0327}_{-0.0320}$ & $0.1128^{+0.0327}_{-0.0139}$ & $84.30^{+1.72}_{-2.05}$ & quad & 4.58	&	7.26	&	-0.30	&	0.48\\
737$^{(4)}$	& $0.1386^{+0.0080}_{-0.0068}$ & $0.1290$ & $83.62$ &  log &                                     3.18	&	23.43	&	-0.18	&	1.39\\
737$^{(3)}$	& $0.1398^{+0.0088}_{-0.0084}$ & $0.1278^{+0.0103}_{-0.0093}$ & $83.62$ & sqrt &                7.24	&	7.13	&	-0.43	&	0.42\\
739	& $0.1446^{+0.0069}_{-0.0073}$ & $0.1282^{+0.0043}_{-0.0037}$ & $83.73^{+0.43}_{-0.38}$ &  log & 0.01	&	1.56	&	0.00	&	0.23\\
758$^{(4)}$	& $0.1487^{+0.0028}_{-0.0055}$ & $0.1248^{+0.0050}_{-0.0047}$ & $83.62$ &  log &               0.24	&	0.19	&	-0.37	&	0.27\\
758$^{(3)}$	& $0.1450$ & $0.1290$ & $83.62$ &  sqrt &                                      1.24	&	0.92	&	-2.05	&	1.53\\
858	& $0.1450$ & $0.1385^{+0.0117}_{-0.0098}$ & $83.78^{+0.16}_{-0.16}$ & sqrt &  0.00	&	4.69	&	0.00	&	0.32\\
971	& $0.1210^{+0.0251}_{-0.0257}$ & $0.1326^{+0.0026}_{-0.0039}$ & $85.39^{+1.89}_{-1.90}$ &  log & -17.05	&	1.32	&	1.28	&	0.10\\
\\ \hline
\end{tabular}\end{small}
\\
\begin{footnotesize}
$^{(a)}$ Limb darkening law used for light curve analysis. For details see text.\\
$^{(b)}$ These two transits were detrended with a linear line because they consist only of $\sim$20 data points, i.e. there are too less information for a second order polynomial fit \end{footnotesize}
\end{table*}

\subsection{Blending}

In \citet{2009AN....330..459R} we already dealt with a faint object close to the TrES-2 host star. Because this object has a projected separation of only $\sim\,10.3$\,arcsec at epoch 2009.7 to the TrES-2 host star we could not resolve both stars independently with our instruments under decent seeing conditions. \\ In addition, \citet{2009A&A...498..567D} detected a very nearby companion candidate (projected separation $1.089\,\pm\,0.009$\,arcsec) with a spectral type of K4.5-K6. Both objects lie within the aperture, which means that the additional light (``third'' light) contaminates the photometry. Simulations by \citet{2010MNRAS.408.1689S} showed that the existence of third light causes an overestimation of $r_{\mathrm{A}}$ and an underestimation of $r_{\mathrm{b}}$ and $i$. If the contribution of the third light ($L_{\mathrm{3}}$) is not considered, systematic errors in the system parameters will arise without changing the quality of the fit.\\ The visual companion detected by \citet{2009A&A...498..567D} is about 30 times fainter than the TrES-2 host star. \citet{2011ApJ...733...36K} were the first who considered the effect of the additional light. Also, \citet{2011MNRAS.417.2166S} and \citet{2012A&A...539A..97S} include the third light in their analysis of the \textit{Kepler} light curves. None of these authors mentioned the faint object shown by \citet{2009AN....330..459R}. An estimate of the brightness of this object, however, shows that the difference in brightness is $\sim$\,6\,mag and thus about 250 times fainter than the TrES-2 host star. The contribution of the additional light is about one order of magnitude less than that of the object detected by \citet{2009A&A...498..567D} and thus negligible. \\ In general, the precision of our ground-based observations of TrES-2 is not sufficient to determine the system parameters with very high accuracy. Since the difference in the system parameters with and without the third light is within our error bars we neglected its contribution. But for the analysis of the high precision \textit{Kepler} light curves (see Section \ref{Kepler_data}) we of course included the contribution of the additional light, which was determined for the wavelength range of \textit{Kepler} by \citet{2011MNRAS.417.2166S} to be $L_{\mathrm{3}}\,=\,0.0258\,\pm\,0.0008$.

\section{\textit{Kepler} data}
\label{Kepler_data}

For our analysis we used \textit{Kepler} data of quarters Q0 to Q17. The data were downloaded from the ``NASA Exoplanet Archive'' \citep[][http://exoplanetarchive.ipac.caltech.edu/]{2013arXiv1307.2944A} which provides fully reduced photometric time series. The used light curves were generated with the 'Kepler Data Processing Pipeline' which includes the reduction, calibration and photometry, as well as the barycentric correction and the detection of systematic effects \citep{2010ApJ...713L..87J,Kepler_Processing}. We downloaded the fully processed ('Presearch Data Conditioning Simple Aperture Photometry', PDCSAP) data, in which systematic errors are already corrected \citep{2010SPIE.7740E..62T}.\\ The \textit{Kepler} light curves cover an observation period from 2009 May 2 to 2013 May 11 and included a total of 435 transits (see Table~\ref{Number_Transits}). TrES-2 was observed in short cadence mode. With a sampling of 58.85\,s covering 1471\,d $\sim$\,1,570,950 data points were collected.\\ During Q4 the CCD chip on which the TrES-2 was observed failed \citep{KDRN6}. Since Kepler is rotated by 90$^{\circ}$ each quarter and thus TrES-2 falls on another chip, it could be observed again from Q5. In consequence of this failure \textit{Kepler} cannot collect data for TrES-2 in every fourth quarter (i.e. no data in Q8, Q12, and Q16). \\ For the analysis of the \textit{Kepler} data we approximated the best-fitting model with \begin{scriptsize}JKTEBOP\end{scriptsize} for each of the 435 individual transits. We again used Table~\ref{Fitparameter_TrES2} for the initial values of the fit parameters. All parameters were set free except of the LD coefficients (see Section \ref{system_parameters}) and the contribution of the third light which was fixed to the value of \citet{2011MNRAS.417.2166S}. We derived the errors by running 1000 Monte Carlo simulations. A confidence range of a given parameter within 68.3\% (1\,$\sigma$) was taken as its error estimate.

\begin{table}
\centering
\caption{\textit{Kepler} observations of TrES-2.}
\label{Number_Transits}
\begin{tabular}{cccc}
\hline \hline
\textit{Q} & Start & End & No. of transits  \\ \hline
0 & 02 May 2009 & 11 May 2009 & 4  \\
1 & 13 May 2009 & 15 Jun. 2009 & 14 \\
2 & 20 Jun. 2009 & 16 Sep. 2009 & 33 \\
3 & 18 Sep. 2009 & 17 Dec. 2009 & 33 \\
4 & 19 Dec. 2009 & 09 Jan. 2010 & 9  \\
5 & 20 Mar. 2010 & 23 Jun. 2010 & 38 \\
6 & 24 Jun. 2010 & 22 Sep. 2010 & 37 \\ 
7 & 23 Sep. 2010 & 22 Dec. 2010 & 35 \\
8 & 06 Jan. 2011 & 14 Mar. 2011 & \\
9 & 21 Mar. 2011 & 26 Jun. 2011 & 39 \\
10 & 27 Jun. 2011 & 28 Sep. 2011 & 38 \\
11 & 29 Sep. 2011 & 04 Jan. 2012 & 37 \\
12 & 05 Jan. 2012 & 28 Mar. 2012 & \\
13 & 29 Mar. 2012 & 27 Jun. 2012 & 35 \\ 
14 & 28 Jun. 2012 & 03 Oct. 2012 & 36 \\
15 & 05 Oct. 2012 & 11 Jan. 2013 & 36 \\
16 & 12 Jan. 2013 & 08 Apr 2013 &  \\
17 & 09 Apr. 2013 & 11 May 2013 & 11 \\\hline
& & & 435 \\ \hline \hline
\end{tabular}
\end{table}

\subsection{Comparison of JKTEBOP and TAP}

Throughout this work the light curves were analysed with \begin{scriptsize}JKTEBOP\end{scriptsize} which estimates the uncertainties either with a simple Monte Carlo simulation or the prayer bead method. An alternative method is to use Markov Chain Monte Carlo (MCMC) to estimate uncertainties. To compare these methods we modelled the 435 Kepler transits light curves again with the Transit Analysis Package\footnote{http://ifa.hawaii.edu/users/zgazak/IfA/\begin{scriptsize}TAP\end{scriptsize}.html} \citep[\begin{scriptsize}TAP\end{scriptsize} v2.1,][]{2012AdAst2012E..30G} using 10 chains with $10^{5}$ steps each. \begin{scriptsize}TAP\end{scriptsize} uses the MCMC techniques to fit transit light curves using the \citet{2002ApJ...580L.171M} model which are calculated using  EXOFAST \citep{2013PASP..125...83E}. To estimate the parameter uncertainties, the wavelet-based technique of \citet{2009ApJ...704...51C} is used to take into account time-correlated noise. It has been shown that this approach provides the most reliable parameters and error estimates \citep[e.g.][]{2012ApJ...748...22H}. \\ The \begin{scriptsize}TAP\end{scriptsize} code employs the quadratic LD law. For a reasonable comparison the modelling with \begin{scriptsize}JKTEBOP\end{scriptsize} (MC and prayer bead) was also done with quadratic LD where the coefficients were kept fixed but their values were perturbed to ensure that this source of error was included in the results. \begin{scriptsize}TAP\end{scriptsize} does not account for third light contribution. Therefore the additional flux from the contaminant was removed from the light curves before the modelling. The results of the comparison is shown in Table~\ref{TAP_vs_JKTEBOP}. For the \textit{Kepler} data \begin{scriptsize}JKTEBOP\end{scriptsize} yield similar error estimates with its two methods except for the transit time which error seems to be underestimated by the residual shift algorithm. The error estimates for \begin{scriptsize}JKTEBOP\end{scriptsize} and \begin{scriptsize}TAP\end{scriptsize} differ by factors of 1.2--1.6, 2.3--2.6, 1.0--1.2, and 1.2--1.4 for $r_{\mathrm{A}}+r_{\mathrm{b}}$, $k$, $i$, and $T_{\mathrm{c}}$, respectively. \\ Although it has been shown that the Monte Carlo or the prayer bead method may lead to underestimated uncertainties, the wavelet-based MCMC technique seems to overestimate the error bars. But this is only true when a small amount ($\sim$zero) of correlated noise is present in the data (space-based-data). For the ground-based data this comparison is maybe not completely correct. More details on this issue can be found in Section \ref{transit_timing}.

\begin{table*}
\centering
\caption{Comparison between JKTEBOPand TAP and their three different error estimation methods.}
\label{TAP_vs_JKTEBOP}
\begin{tabular}{cccc}
\hline \hline
Parameter & \begin{scriptsize}JKTEBOP\end{scriptsize} & \begin{scriptsize}JKTEBOP\end{scriptsize}& \begin{scriptsize}TAP\end{scriptsize} \\ 
& MC & prayer bead & MCMC + wavelet method \\\hline
$r_{\mathrm{A}}+r_{\mathrm{b}}$* &  $0.1435^{+0.0010}_{-0.0011}$ & $0.1435^{+0.0010}_{-0.0011}$ & $0.1427^{+0.0016}_{-0.0013}$\\
$k$ & $0.1273^{+0.0007}_{-0.0008}$ & $0.1273^{+0.0006}_{-0.0007}$ & $0.1272^{+0.0018}_{-0.0016}$ \\
$i$  [$^{\circ}$] & $83.80^{+0.10}_{-0.09}$ & $83.80^{+0.11}_{-0.10}$ &  $83.86^{+0.12}_{-0.10}$ \\ 
$T_{\mathrm{c}}$ - 2454955  [d] & $0.762500^{+0.000050}_{-0.000051}$ & $0.762496^{+0.000049}_{-0.000037}$ & $0.762517^{+0.000063}_{-0.000061}$ \\
\hline \hline
\end{tabular}
\\
$^{\ast}$This parameter is not fitted by the \begin{scriptsize}TAP\end{scriptsize} algorithm but was converted from the two output parameters $k$ and $a/R_{\mathrm{A}}$ 
\end{table*} 

\subsection{Limb darkening}

\begin{figure}
  \centering
  \includegraphics[width=0.33\textwidth,angle=270]{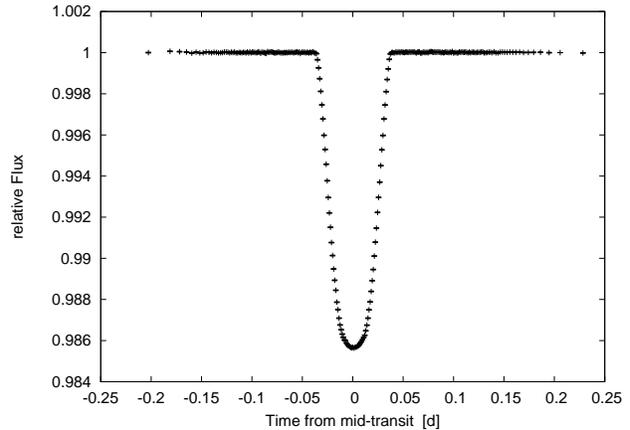}
  \caption{84 individual \textit{Kepler} light curves were used to create this phase-folded and binned light curve of TrES-2. Always 125 measurements were combined into one data point.}
  \label{Lc_TrES2_Flux}
\end{figure}

\begin{figure}
\begin{minipage}[]{0.45\textwidth}
  \centering
  \includegraphics[height=0.23\textheight]{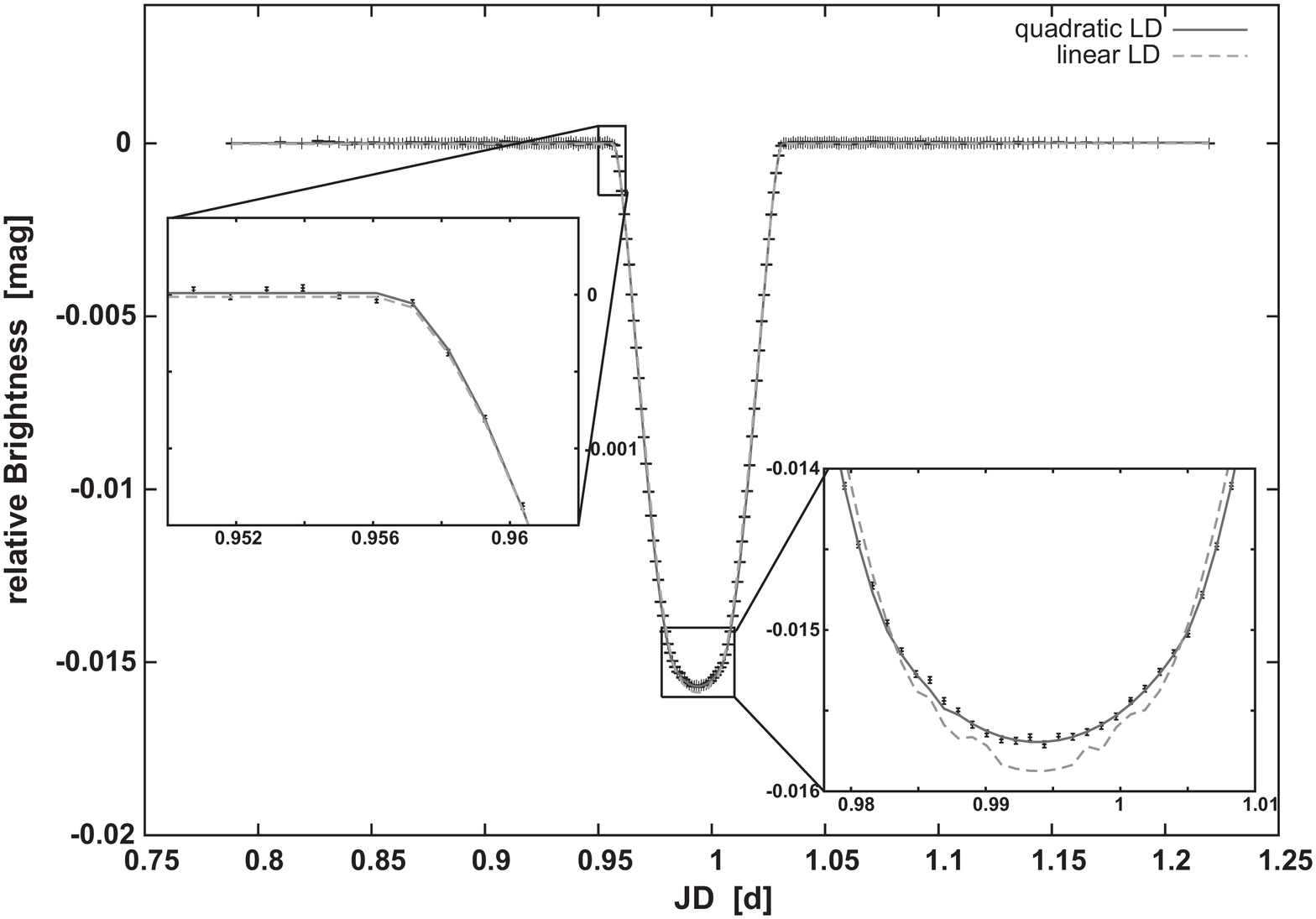}
  \caption{The phased-folded and binned light curve from Fig. \ref{Lc_TrES2_Flux} modelled with the quadratic (solid black line) and the linear (dashed grey line) LD law.}
  \label{TrES2_LD_Modelle}
\end{minipage}
\begin{minipage}[]{0.45\textwidth}
  \centering
  \includegraphics[height=0.34\textheight, angle=270]{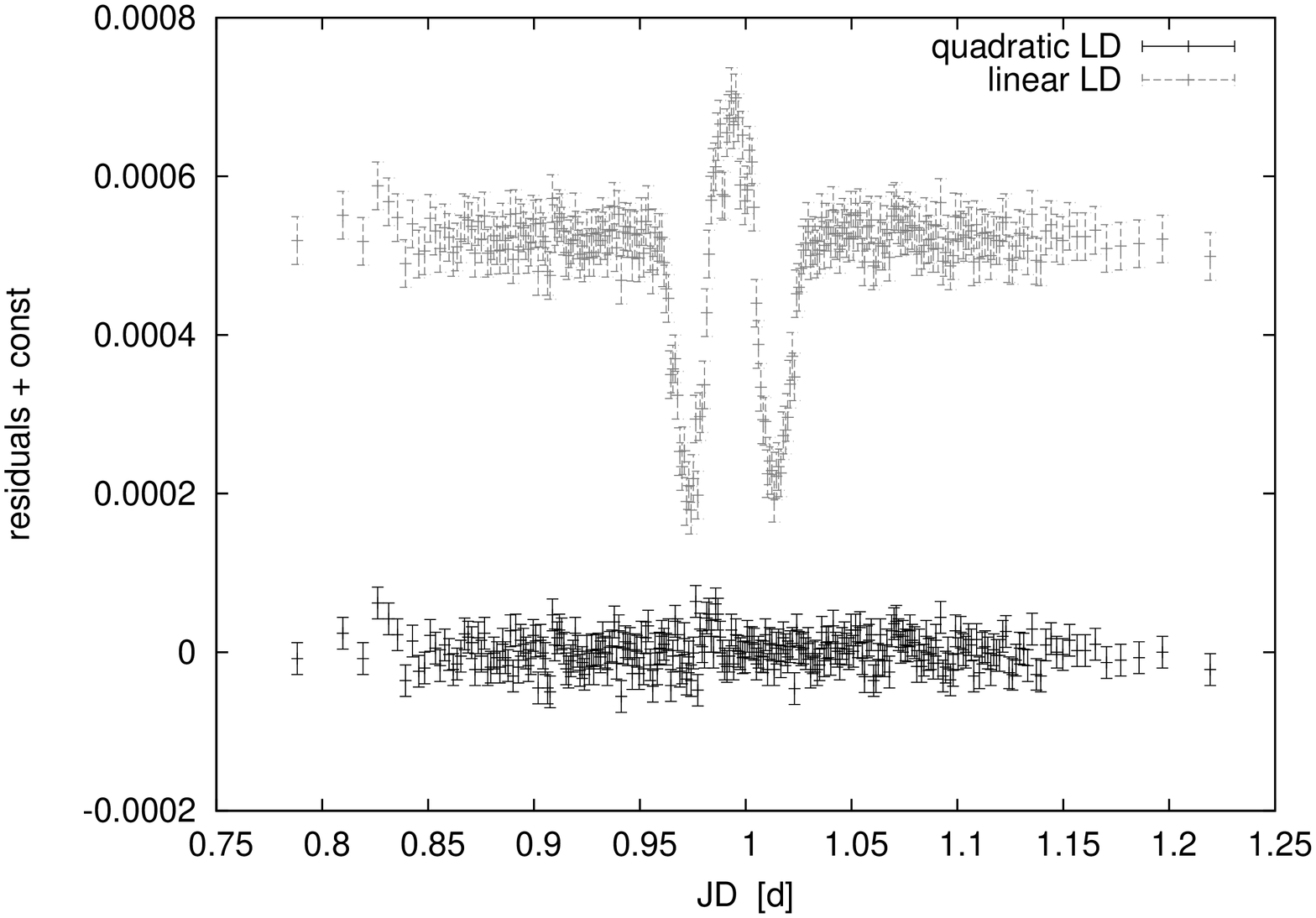}
  \caption{Residuals between data points and the model from Fig. \ref{TrES2_LD_Modelle}. The quadratic and the linear case are shown as black solid and grey dotted points, respectively.}
  \label{Residuen}
\end{minipage}
\end{figure}

To determine the effect of the LD and its coefficients, 84 transits (Q0--Q3) were phase-folded and binned. After applying a 2$\sigma$-clipping always 125 points were combined into one data point. Since we already reached a S/N\,$\sim$\,760 which is sufficient to derive LD coefficients \citep{2013A&A...549A...9C} we did not repeat the phase-folding with all 18 quarters. The resulting light curve is shown in Fig.~\ref{Lc_TrES2_Flux}. \\ Because of the very high S/N  we could compare the observed LD coefficients with the theoretical predictions in the tables of \citet{2010A&A...510A..21S}. The analysis was done with \begin{scriptsize}JKTEBOP\end{scriptsize} with the quadratic LD law and the initial values from Table~\ref{Fitparameter_TrES2}. While fitting, most parameters were allowed to vary, only the LD coefficient were fixed to the values from Table~\ref{LD_TrES2}. The analysis was repeated with the linear law and a fixed LD coefficient of $u_{\mathrm{lin}}=0.558$ \citep{2010A&A...510A..21S}. All parameters were fixed to the values of the previous fit except for the transit depth which depends on the LD. Fig.~\ref{TrES2_LD_Modelle} shows the difference between the two approaches, Fig.~\ref{Residuen} shows the residuals for clarity. The linear LD overestimates the depth of the transit while the quadratic law reproduces the observations. The deviations from a smooth curve in the fit suggest numerical errors of the model which is expected around the transit centre for almost all models based on numerical integration over the visible surface of the components as a result of the very different radii of the two components \citep{2013MNRAS.431.3654K}. Quantitatively, modelling the light curve with the quadratic LD yields an rms of 0.02\,mmag while for the linear LD law the rms is 0.09\,mmag, 4.5 times higher. In summary, the LD coefficients in the tables of \citet{2010A&A...510A..21S} for the \textit{Kepler} wavelength reproduces the observations if using the quadratic LD law but they fail if using the linear LD law. The analysis confirms the results of \citet{2007A&A...467.1215S}. If the LD coefficients for the quadratic law were allowed to vary during the analysis the result is within the error bars consistent with the values of \citet{2010A&A...510A..21S} (see Table~\ref{Systemparameter_TrES2}, first column).

\subsection{System parameters}
\label{system_parameters}

Since the fit with a linear LD law yields worse results, the system parameters were derived from the fit with quadratic LD. The result is given in Table~\ref{Systemparameter_TrES2} (first column). The values of the parameters for quadratic LD are consistent with the results of \citet{2011ApJ...733...36K}, \citet{2011MNRAS.417.2166S} and \citet{2012A&A...539A..97S}. The derived LD coefficients are used for the light curve analysis of the 435 transits.\\ Since the modelling of the phase-folded and binned light curve did not improve the results much with respect to previous works we tried a different approach. With \begin{scriptsize}TAP\end{scriptsize} it is possible to fit several transits simultaneously. Because of this large number of 435 \textit{Kepler} transits the simultaneous modelling was done stepwise. Each observational quarter is divided in three parts. All transits of each part were modelled and then combined which gave a number of $\sim$40 combined transits. In the second step we fitted these combined transits simultaneously. The final results are given in Table~\ref{Systemparameter_TrES2} (second column). Our results using two different approaches are entirely consistent but the simultaneous modelling yields up to 6.5 times more precise values. Compared to the results of \citet{2011MNRAS.417.2166S} our results are even more precise by up to a factor of 9.

\begin{table*}
\centering
\caption{System parameters derived from the light curve analysis of the phase-folded and binned \textit{Kepler} light curve and from the (quasi)-simultaneous analysis with TAP in comparison to \citet{2011MNRAS.417.2166S}.}
\label{Systemparameter_TrES2}
\renewcommand{\arraystretch}{1.2} 
\begin{tabular}{cccc}
\hline \hline
Parameter & \multicolumn{2}{c}{This work} & \citet{2011MNRAS.417.2166S}$^{a}$\\ 
& Average LC & Simultaneous fit & \\ \hline
$r_{\mathrm{A}}+r_{\mathrm{b}}$ &  $0.14181\pm 0.00054$ & $0.14206\pm 0.00012^{b}$ & $0.14097\pm 0.00089$\\
$k$ & $0.12654\pm 0.00097$ & $0.12656\pm 0.00015$ & $0.12817 \pm 0.00139$ \\
$i$  [$^{\circ}$] & $83.905\pm 0.040$ & $83.908\pm 0.009$ & $83.970 \pm 0.063$ \\
$u_{\mathrm{q}}$ & $0.35\pm0.14$ & $0.26\pm 0.03$ & $0.56 \pm 0.21$ \\
$v_{\mathrm{q}}$ & $0.23\pm0.15$ & $0.35\pm 0.03$ & $0.03 \pm 0.22$ \\
$r_{\mathrm{A}}$ & $0.12588\pm 0.00059$ & $0.12610\pm 0.00011^{b}$ & $0.12495 \pm 0.00095$ \\
$r_{\mathrm{b}}$ & $0.01593\pm 0.00006$ & $0.01596\pm 0.00002^{b}$ & $0.01602 \pm 0.00006$ \\
\hline \hline
\end{tabular}
\\
$^{a}$Derived from analysis with a quadratic LD law and free LD coefficients of a phase-folded and binned \textit{Kepler} light curve consisting of 52 transits (Q0--Q2). \\
$^{b}$This parameter is not fitted by the \begin{scriptsize}TAP\end{scriptsize} algorithm but was calculated from the two output parameters $k$ and $a/R_{\mathrm{A}}$. 
\end{table*}

\subsection{Physical properties}

The results shown in Table~\ref{Systemparameter_TrES2} allow us to calculate stellar, planetary and geometrical parameters. We used the values that were obtained by the simulations modelling using the \begin{scriptsize}TAP\end{scriptsize} algorithm (column 2 in Table~\ref{Systemparameter_TrES2}). By inserting the improved period (see the next section) into Kepler's third law we derived the semi-major axis $a$. The necessary masses were taken from Table~\ref{Werte_TrES2}. We then translated the fractional radii into a true radius for star $R_{\mathrm{A}}$ and planet $R_{\mathrm{b}}$ and calculated the stellar and planetary density $\rho_{\mathrm{A}}$ and $\rho_{\mathrm{b}}$. From the stellar radius $R_{\mathrm{A}}$, the semi-major axis $a$ and the orbital inclination $i$ we calculated the impact parameter $b$. Simplified formulas for determining the surface gravity can be found in \citet{2009MNRAS.394..272S}. For this purpose the radial velocity amplitude of \citet{2006ApJ...651L..61O} has been used. The eccentricity was set to zero.\\ The equilibrium temperature of the planet, $T_{\mathrm{eq}}$, was derived assuming the effective temperature of the host star from Table~\ref{Werte_TrES2} and using the relation in \citet{2007ApJ...671..861H}. In addition we calculated the Safronov number $\Theta$ \citep{1972epcf.book.....S}, which is proportional to the ratio of the escape velocity of the planet $v_{\mathrm{esc}}$ and the orbital velocity $v_{\mathrm{orb}}$ \citep{2007ApJ...671..861H}. Basically it measures the efficiency with which a planet gravitationally scatters other bodies. This information plays an important role in the understanding of migration.\\ The results of the calculations are summarized in Table~\ref{phys_prop_TrES2}. Within the error bars all parameters are in agreement with the literature values. Although the system parameters could be determined with a very high precision, the results of the physical properties are not much better than the values in previous works. This is because of the usage of the literature values for the masses of star and planet as well as the temperature and the radial velocity amplitude which limits the final precision.

\begin{table}
\centering
\caption[]{Physical properties of the TrES-2 system derived from light curve modelling. Literature values given for comparison were derived by \citet{2011MNRAS.417.2166S} except of the impact parameter $b$ which was taken from \citet{2011ApJ...733...36K}.}
\label{phys_prop_TrES2}
\begin{tabular}{lr@{\,$\pm$\,}lr@{\,$\pm$\,}l}
\hline \hline
 Parameter & \multicolumn{2}{c}{This work} & \multicolumn{2}{c}{Literature values} \\ \hline \hline
& \multicolumn{4}{c}{Planetary parameters} \\ \hline 
$R_{\mathrm{b}}$  [R$_{\mathrm{Jup}}$] & 1.189 & 0.025 & 1.193 & 0.021\\
$\rho_{\mathrm{b}}$  [$\mathrm{\rho}_{\mathrm{Jup}}$] & 0.703 & 0.053 & 0.665 & 0.015 \\
$g_{\mathrm{b}}$  [$\frac{m}{s^{2}}$] & 21.07 & 0.31 & 21.02 & 0.31 \\
$T_{eq}$ [K] & 1455 & 18 & 1466 & 12 \\ 
$\Theta$ & 0.0769 & 0.0058 & 0.0727 & 0.0017 \\\hline
& \multicolumn{4}{c}{Stellar parameters} \\ \hline 
$R_{\mathrm{A}}$  [R$_{\mathrm{\odot}}$] & 0.963 & 0.020 & 0.964 & 0.017\\
$\rho_{\mathrm{A}}$  [$\mathrm{\rho}_{\mathrm{\odot}}$] & 1.095 & 0.098 & 1.105 & 0.011 \\ 
log\,$g_{\mathrm{A}}$ & 4.461 & 0.009 & 4.466 & 0.008 \\\hline
& \multicolumn{4}{c}{Geometrical parameters} \\ \hline 
$a$  [AE] & 0.03555 & 0.00075 & 0.03567 & 0.00061 \\
$i$  [$^{\circ}$] & 83.908 & 0.009 & 83.925 & 0.030 \\
$b$ & 0.842 & 0.001 & 0.848 & $^{0.022}_{0.018}$ \\ \hline \hline
\end{tabular}
\end{table}

\section{Variation in system parameters}

\begin{figure}
\begin{minipage}[]{0.45\textwidth}
  \centering
  \includegraphics[height=0.32\textheight, angle=270]{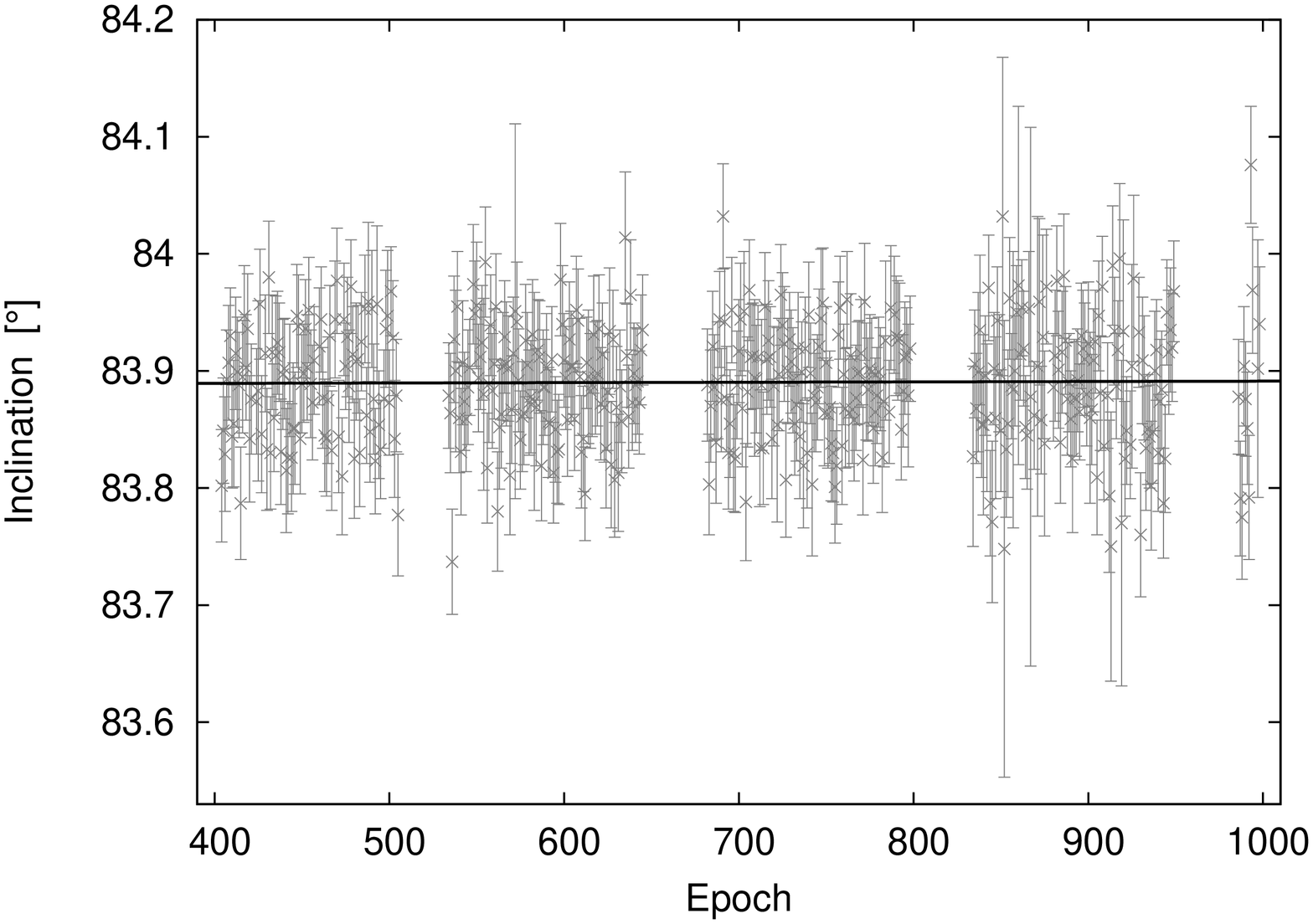}
  \caption{Inclination $i$ versus epoch for all 435 \textit{Kepler} transits. The black line represents the best weighted linear fit.}
  \label{Epoch_Inkl}
\end{minipage}
\begin{minipage}[]{0.45\textwidth}
  \centering
  \includegraphics[height=0.34\textheight, angle=270]{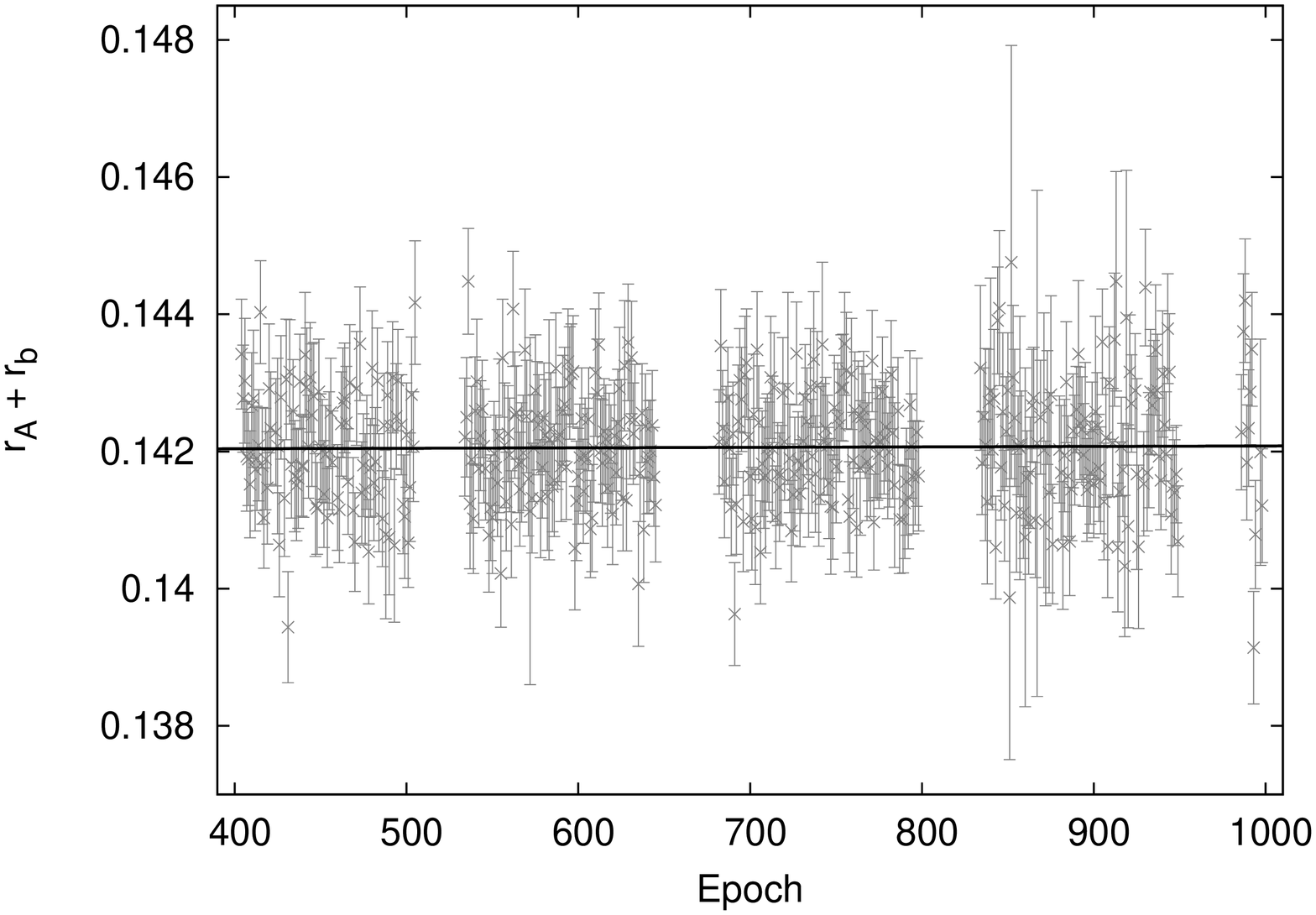}
  \caption{As Fig.~\ref{Epoch_Inkl} but for the sum of the fractional radii $r_{\mathrm{A}}+r_{\mathrm{b}}$.}
  \label{Epoch_r1pr2}
\end{minipage}
\begin{minipage}[]{0.45\textwidth}
  \centering
  \includegraphics[height=0.34\textheight, angle=270]{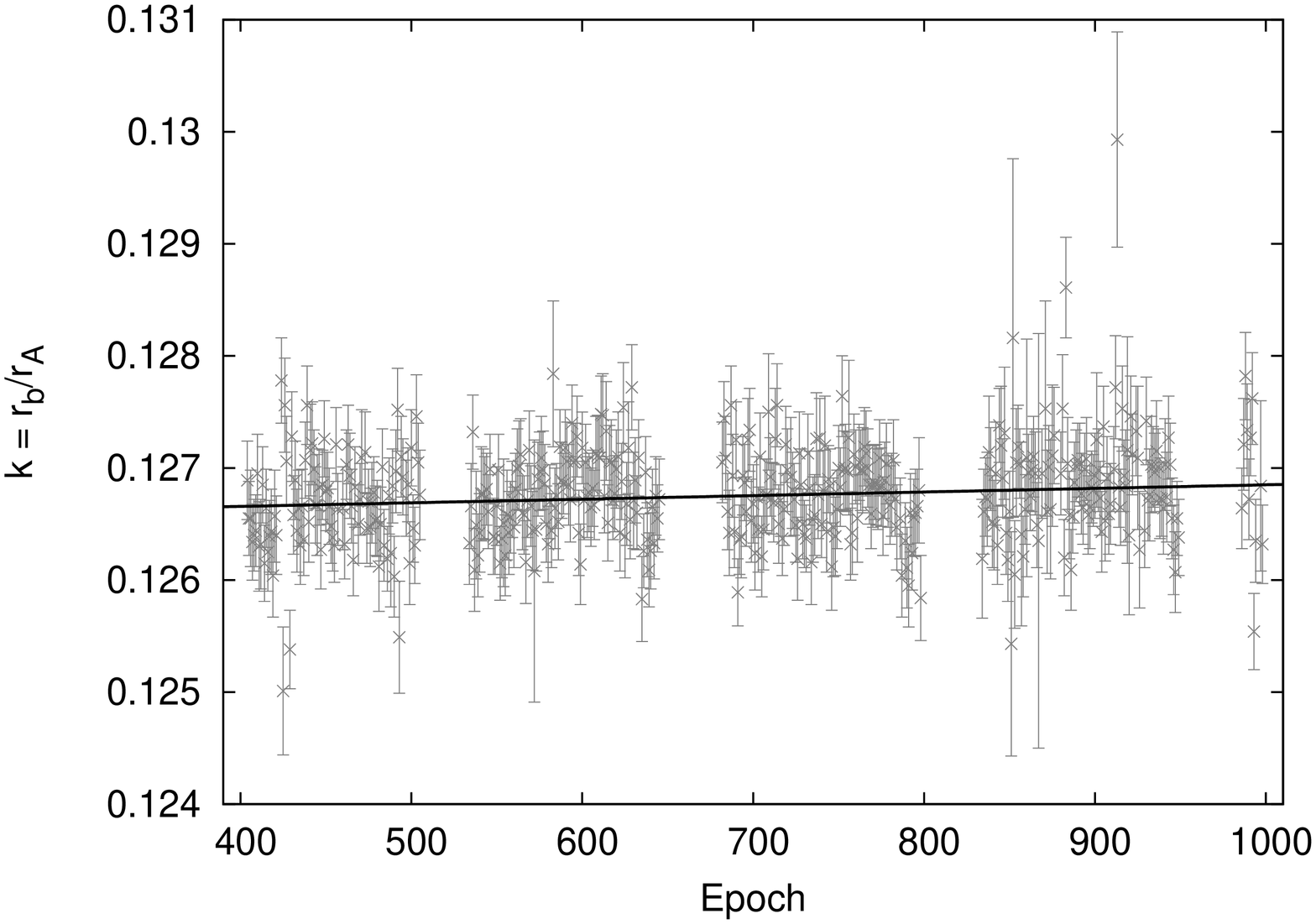}
  \caption{As Fig.~\ref{Epoch_Inkl} but for the radius ratio $k$.}
  \label{Epoch_k}
\end{minipage}
\end{figure}

Owing to the high S/N, with the observations of \textit{Kepler} it is possible to determine the system parameters independently for each transit. Thus, any variation can be detected. Figs.~\ref{Epoch_Inkl}--\ref{Epoch_k} show $i$, $r_{\mathrm{A}}+r_{\mathrm{b}}$ and $k$ versus epoch. A summery of the parameters for all 435 \textit{Kepler} transits can be found in Table \ref{Results_LCA} which is presented in its entiretly in the online version of this article.\\ For each parameter, we did a weighted linear regression. While the analysis of \citet{2012A&A...539A..97S} yields a slight inclination increase of \mbox{$\Delta i=(8 \pm 2)\times 10^{-5}$ $^{\circ}$ per epoch}, we could not confirm their result. With our larger data set we found a value of \mbox{$\Delta i=(0 \pm 1)\times 10^{-5}$ $^{\circ}$ per epoch} which is consistent with zero. The rms of our weighted linear fit is 0.05$^{\circ}$ and the $\chi^{2}=428.9$. The fitting of a constant inclination is satisfactory with reduced $\chi^{2}$ close to 1, leaving no space for any other more complicated model. To compare our finding to the result of \citet{2012A&A...539A..97S} we fitted their slope to our data. The resulting fit is slightly worse with $\chi^{2}=499.6$. To test if we can reproduce the results of \citet{2012A&A...539A..97S} with our method we repeated the weighted linear fit with the same 84 \textit{Kepler} transits they used (Q0--Q3). We found \mbox{$\Delta i=(16 \pm 19)\times 10^{-5}$ $^{\circ}$ per epoch} which is within our more conservative error bar consistent with zero but also consistent with the result of \citet{2012A&A...539A..97S}. If we use only the first 84 transits we cannot confirm the reported slope but we also cannot completely rule out the result of \citet{2012A&A...539A..97S}. But the larger timespan of 4 years gives tighter constraints on any variation. Using all the \textit{Kepler}-transits for the fitting we found no evidence of an increasing inclination.\\ For the sum of the fractional radii our analysis yields similar results as for the orbital inclination. The fit with an rms of 8.6$\times10^{-4}$ gave a value of \mbox{$\Delta(r_{\mathrm{A}}+r_{\mathrm{b}})=(1\pm 2)\times10^{-7}$ 1/epoch}, meaning that no deviation from zero is noticeable. The transit duration $D$ which was calculated from the orbital period, the inclination and the sum of the fractional radii also does not show any variation (\mbox{$\Delta D=(1 \pm 4)\times 10^{-7}$ d perepoch}). \\ Only the radius ratio, hence the transit depth, shows an increase of \mbox{$\Delta k=(3.2\pm1.0)\times10^{-7}$ 1/epoch} (rms of the fit  4.4$\times10^{-4}$). This result is significant with 3.2\,$\sigma$. To test if we can reproduce this result we repeated the calculation after removing all points that deviate more than 2\,$\sigma$ from the average of the $k$-values for all \textit{Kepler}-transits. The weighted linear fit of the remaining 384 points yielded a slightly lower slope of \mbox{$\Delta k=(2.7\pm1.1)\times10^{-7}$ 1/epoch} which is still significant with 2.5\,$\sigma$. Since the $k$ values are a result of the light curve analysis with \begin{scriptsize}JKTEBOP\end{scriptsize} and a simple MC simulation it is likely that the uncertainties are underestimated. After rescaling the error bars with 1.08 (result only from \textit{Kepler}-data; see Section \ref{transit_timing}) or 1.25 (maximum value of average $\beta$ factor for ground- and space-based observations; see Section \ref{transit_timing}) the positive slope is still significant with 2.9\,$\sigma$ and 2.4\,$\sigma$, respectively. Therefore a slight increase of the radius ratio seems to be evident. \\ A possible explanation could be a change in the third light contamination which was kept fixed during the analysis. However, any change in the amount of additional light coming from a contaminant would not only affect the radius ratio but also change the sum of the radii and the inclination \citep{2010MNRAS.408.1689S}. Since this is not the case, as shown above, the variability of the companion candidate can be ruled out. \\ An increase of the radius ratio $k$ means either an increase of the fractional planetary radius $r_{\mathrm{b}}$ or a decrease of the fractional stellar radius $r_{\mathrm{A}}$. To test these hypotheses, the radii of star and planet were considered individually. Both $r_{\mathrm{A}}$ and $r_{\mathrm{b}}$ show no significant deviation from a constant value. Since $k$ and $r_{\mathrm{A}}+r_{\mathrm{b}}$ are directly fitted values and $r_{\mathrm{A}}$ and $r_{\mathrm{b}}$ are just derived from these parameters, any deviation may be lost in noise. In general, $k$ is about two times more precise than $r_{\mathrm{A}}+r_{\mathrm{b}}$ i.e. any variation will be first visible in the radius ratio. \\ As we do not know a mechanism that could drive the increase of fractional planetary radius so fast (within the 4 years of \textit{Kepler} data), it is more likely that stellar activity i.e. stellar spots could be responsible for changes of the average temperature, hence the effective stellar radius. To evaluate this scenario we estimated the spotted area on the TrES-2 host star. The positive trend in $k$ could be translated into a decreasing luminosity of 0.28\% within the observation period. This was transferred into a spot covered area assuming a constant stellar radius and a spot temperature of 4500K (compared to the temperature of 5795K of the G0V star). These calculations resulted in a change of the spot coverage of 0.44\%. Note that no spot signature could be found in the \textit{Kepler} transit light curves. \\ To check if the values of the spot coverage are realistic we compared it to the 11\,yr solar cycle. \citet{2006Natur.443..161F} determined the variation amplitude of the solar constant within the last three solar cycles to $\sim0.90\,Wm^{-2}$ with an average minimum of $1365.52\,\pm\,0.009\,Wm^{-2}$. If translated to luminosity this yields a variation of 0.066\%. This value is four times smaller than the one of the TrES-2 host star. Note that the last three solar cycles lie within the modern maximum. If we also consider the Maunder Minimum, a period of greatly reduced sunspot activity in the years 1645--1715, the luminosity variation can be as high as 0.2\% \citep{2000GeoRL..27.2425L} which is comparable to the value of the TrES-2 host star. The luminosity variation of the Sun was translated into a spot covered area of 0.1\% (0.34\% including the Maunder Minimum). However, the Sun is known to be a magnetic very inactive star, i.e. it is near the bottom of the distribution of variability of solar-like stars \citep{1998ApJS..118..239R}. This means that it is likely that the TrES-2 host star could have a higher spot covered area. \\  From spectroscopic observations and measurements of the chromospheric activity from the Ca\,II H and K lines \citet{2007ApJ...664.1190S} found a chromospheric emission ratio of \mbox{$\langle log\,R'_{HK}\rangle\,=\,-5.16\,\pm\,0.15$}. This value implies a very low activity of the TrES-2 host star which would correspond to a very high age of $t\sim8$\,Gyr. The TrES-2 host star shows an unusual high Li abundance as it is expected from a star at the age of 1-2\,Gyr. \citet{2007ApJ...664.1190S} speculated on the possibility of a self-enrichment and stated that the age of 1-2\,Gyr is just a lower limit. By comparing the directly fitted parameters from light curve analysis, that are related to the stellar density, and the effective temperature with stellar evolution models \citet{2007ApJ...664.1190S} gave a final age of $5.1^{+2.7}_{-2.3}$\,Gyr. The results imply that the TrES-2 host star is slightly older than the Sun and therefore less active. However, the spectra used in \citet{2007ApJ...664.1190S} were taken in 2006 and would not disagree with a changing stellar activity. The positive trend in the radius ratio indicates an increasing activity. Hence, it is likely that the activity level in 2006 was low. The Ca\,II H \& K activity index for the TrES-2 host star of \mbox{$\langle log\,R'_{HK}\rangle\,=\,-4.949$} found by \citet{2010ApJ...720.1569K} also strengthens the claim of an rising stellar activity. \\ The increasing activity of the host star should also influence the out-of-transit light curve. Over the observation period of 4 years the brightness of the star should show a negative long-term trend. We could not find the brightness decrease of 0.3\% that is expected from the luminosity change. \citet{2012ApJ...761...53B} found in the light curve evidence for ellipsoidal variation, Doppler beaming and the reflection effect also did not mention any long-term trends. However, the data they used have been normalized to the median for each quarter individually. Hence any trend on time-scales longer than 90\,d has been removed. The instrumental setup (the rotation of \textit{Kepler} in every quarter) and the consequential systematics make it impossible to find the trend on the expected level in the unnormalized light curve. As there is no sign of stellar activity and rotation in the light curve, the long-term changes in spot coverage seem to be a plausible explanation.\\ \citet{2012A&A...548A..88A} investigated the Sun’s 11 year activity cycle, which is modulated on longer time-scales (e.g. the 88\,yr Gleissberg and the 208\,yr de Vries cycles). They computed the time-dependent planetary influence on a non-spherical tachocline of the Sun. In this manner \citet{2012A&A...548A..88A} could reproduce long-term cycles of the solar activity. Although \citet{2013A&A...557A..83C} showed that there is no statistically significant evidence for an effect of the planets on solar activity, it has long been speculated  that close-in exoplanets can influence the stellar activity level \citep[e.g.][]{2014A&A...565L...1P}. The long-term magnetic activity of the TrES-2 host star could be modulated by planetary effects. Since TrES-2 is a hot Jupiter, its host star could reveal activity cycles on different time-scales and with different amplitudes. To test the hypothesis it is necessary to spectroscopically and photometrically monitor TrES-2 and its host star for the next years to decades.

\begin{table*}
\caption{Results of the light curve analysis for all \textit{Kepler}-Transits. Table 12 is presented in its entiretly as online-only material. A portion is shown here for guidance regarding its form and content. The uncertainties of the mid-transit times were rescaled with a factor of 1.25 (see text).}
\label{Results_LCA}
\begin{tabular}{cccccc}
\hline \hline
No. & $r_{\mathrm{A}}+r_{\mathrm{b}}$ & $k$ & $i$ [$^{\circ}$] & $T_{\mathrm{c}}$ [BJD$_{\mathrm{UTC}}$] & $\sigma$ [mmag] \\ \hline
1	& $0.14342^{+0.00080}_{-0.00077}$ & $0.12689^{+0.00035}_{-0.00032}$ & $83.802^{+0.048}_{-0.048}$ & $2454955.762505^{+0.000060}_{-0.000062}$ &	0.259	\\
2	& $0.14277^{+0.00078}_{-0.00079}$ & $0.12655^{+0.00033}_{-0.00034}$ & $83.849^{+0.045}_{-0.048}$ & $2454958.233201^{+0.000059}_{-0.000063}$ &	0.255	\\
3	& $0.14303^{+0.00091}_{-0.00081}$ & $0.12655^{+0.00043}_{-0.00037}$ & $83.829^{+0.049}_{-0.055}$ & $2454960.703759^{+0.000064}_{-0.000073}$ &	0.285	\\
4	& $0.14193^{+0.00080}_{-0.00071}$ & $0.12634^{+0.00034}_{-0.00031}$ & $83.893^{+0.042}_{-0.046}$ & $2454963.174384^{+0.000064}_{-0.000065}$ &	0.257	\\
5	& $0.14189^{+0.00068}_{-0.00085}$ & $0.12640^{+0.00032}_{-0.00033}$ & $83.907^{+0.049}_{-0.043}$ & $2454965.644940^{+0.000059}_{-0.000058}$ &	0.255	\\
... & ... & ... & ... & ... & ... \\
\hline\hline
\end{tabular}
\end{table*}

\section{Transit timing}
\label{transit_timing}
 
\begin{figure*}
\begin{minipage}[]{0.48\textwidth}
  \centering
  \includegraphics[height=0.32\textheight, angle=270]{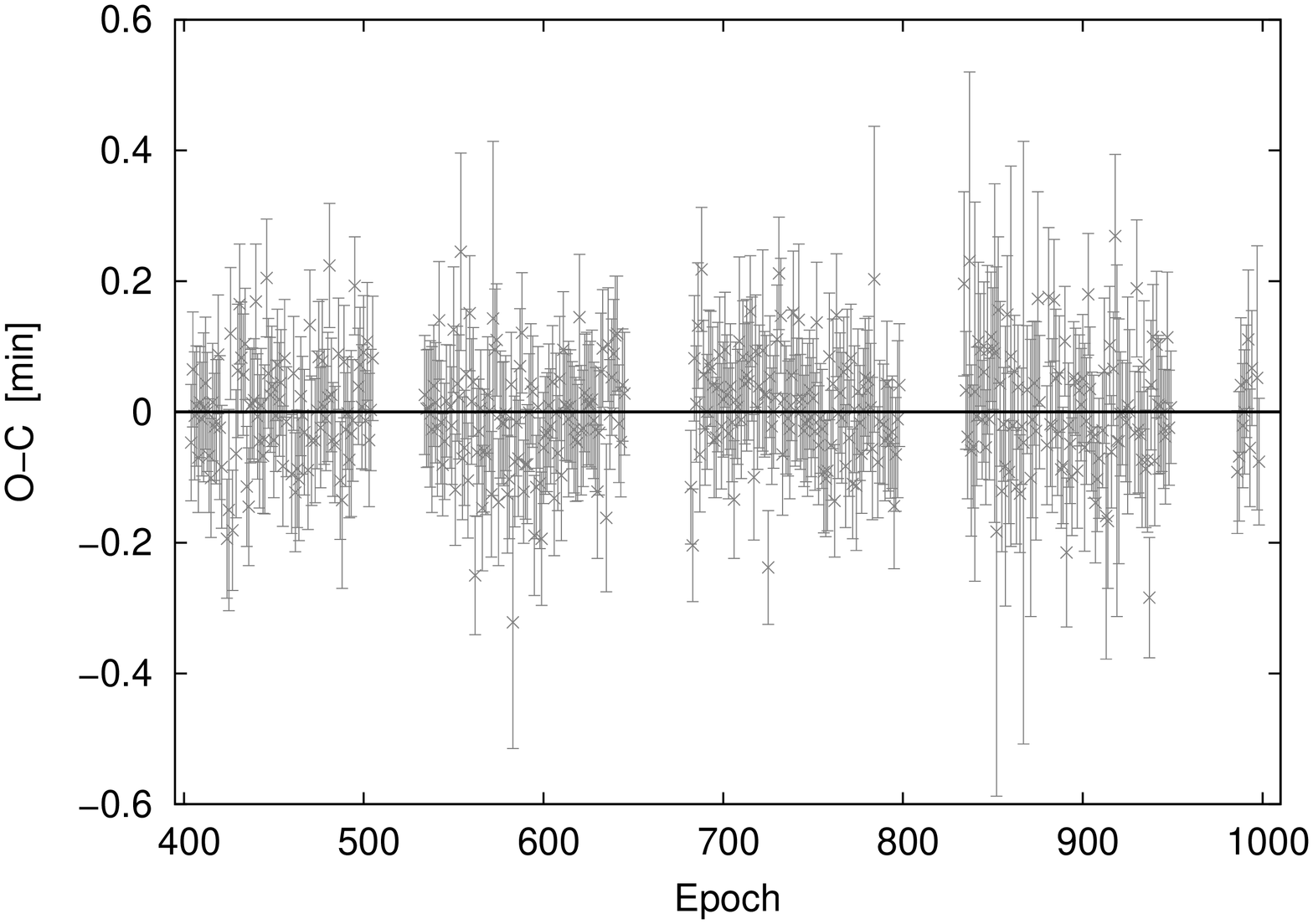}
  \caption{The O--C diagram for the Kepler data, calculated with respect to the updated ephemeris (equation \ref{Elemente_TrES2}).}
  \label{O_C_Kepler}
\end{minipage}
\hspace{0.2cm}
\begin{minipage}[]{0.48\textwidth}
  \centering
  \includegraphics[height=0.32\textheight, angle=270]{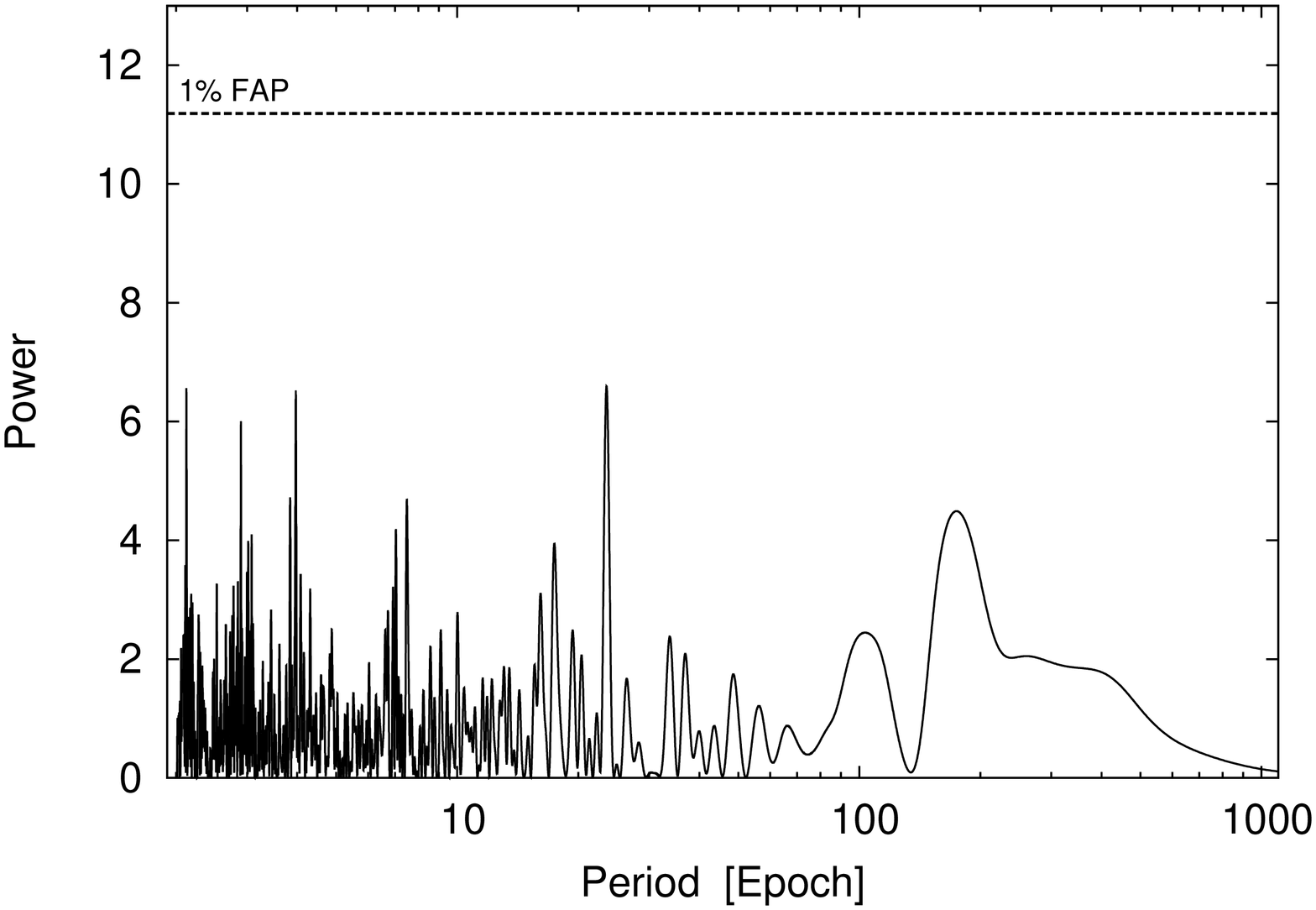}
  \caption{Periodogram of the O--C values of the \textit{Kepler} transits. The dashed vertical line shows the empirical FAP level of 1\%.}
  \label{O_C_Kepler_Fourier}
\end{minipage}
\end{figure*}

\begin{figure}
\begin{minipage}[]{0.45\textwidth}
\centering
  \includegraphics[width=0.7\textwidth, angle=270]{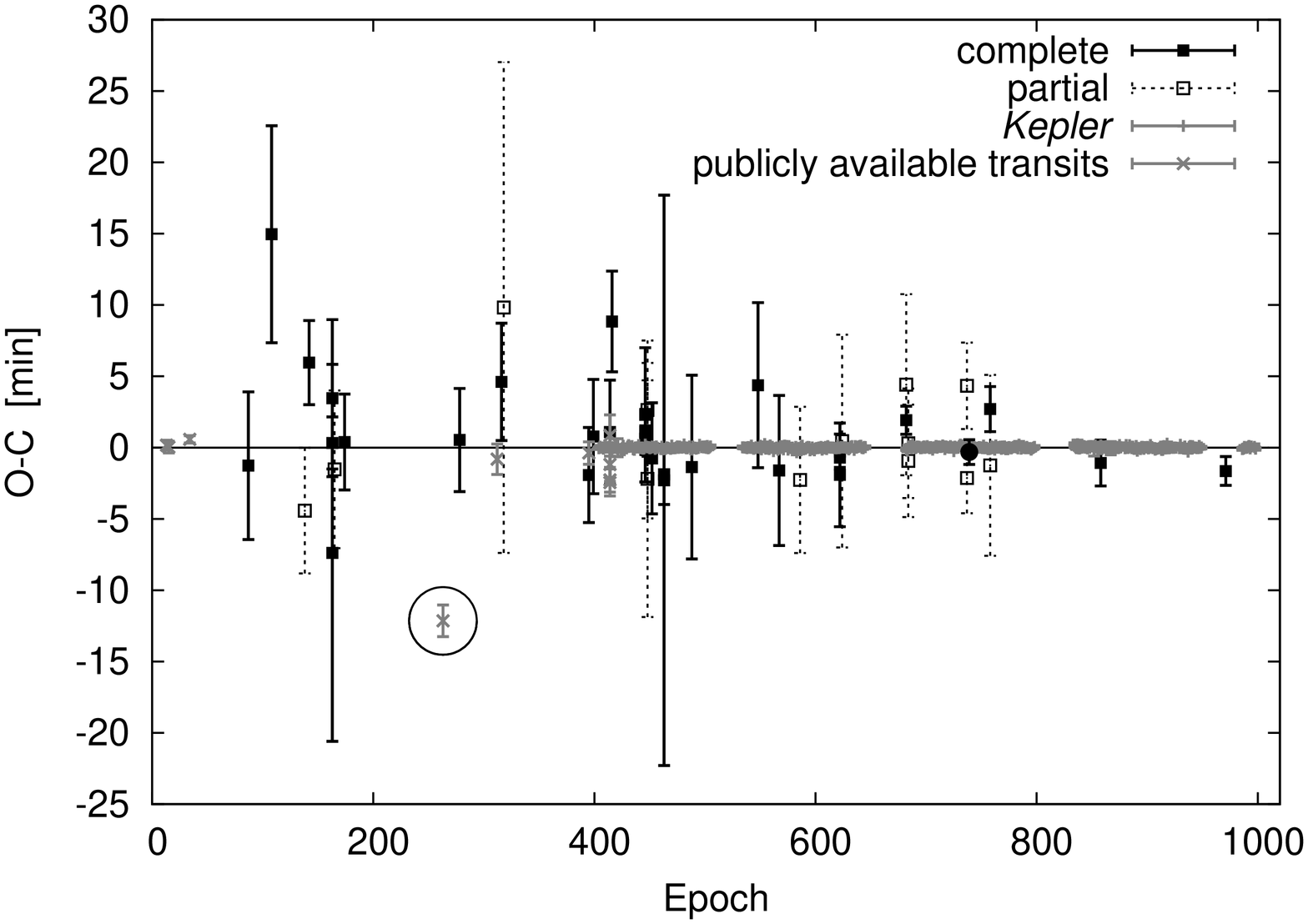}
  \caption{Final O--C diagram: our ground-based observations combined with all 435 \textit{Kepler} transits. The black filled and open (with dashed error bars) symbols denote the complete and the partial ground-based transits, respectively. The \textit{Kepler} and publicly available transit observations are shown in grey. Our best ground-based observation obtained with the Calar Alto 2.2-m telescope is marked by a black filled circle. The solid line represents the updated ephemeris given in equation \ref{Elemente_TrES2}. The circle marks the first transit of \citet{2009A&A...500L..45M} that deviates from the zero line by 11$\sigma$.}
  \label{O_C_all}
\end{minipage}
\begin{minipage}[]{0.45\textwidth}
  \centering
  \includegraphics[width=0.7\textwidth, angle=270]{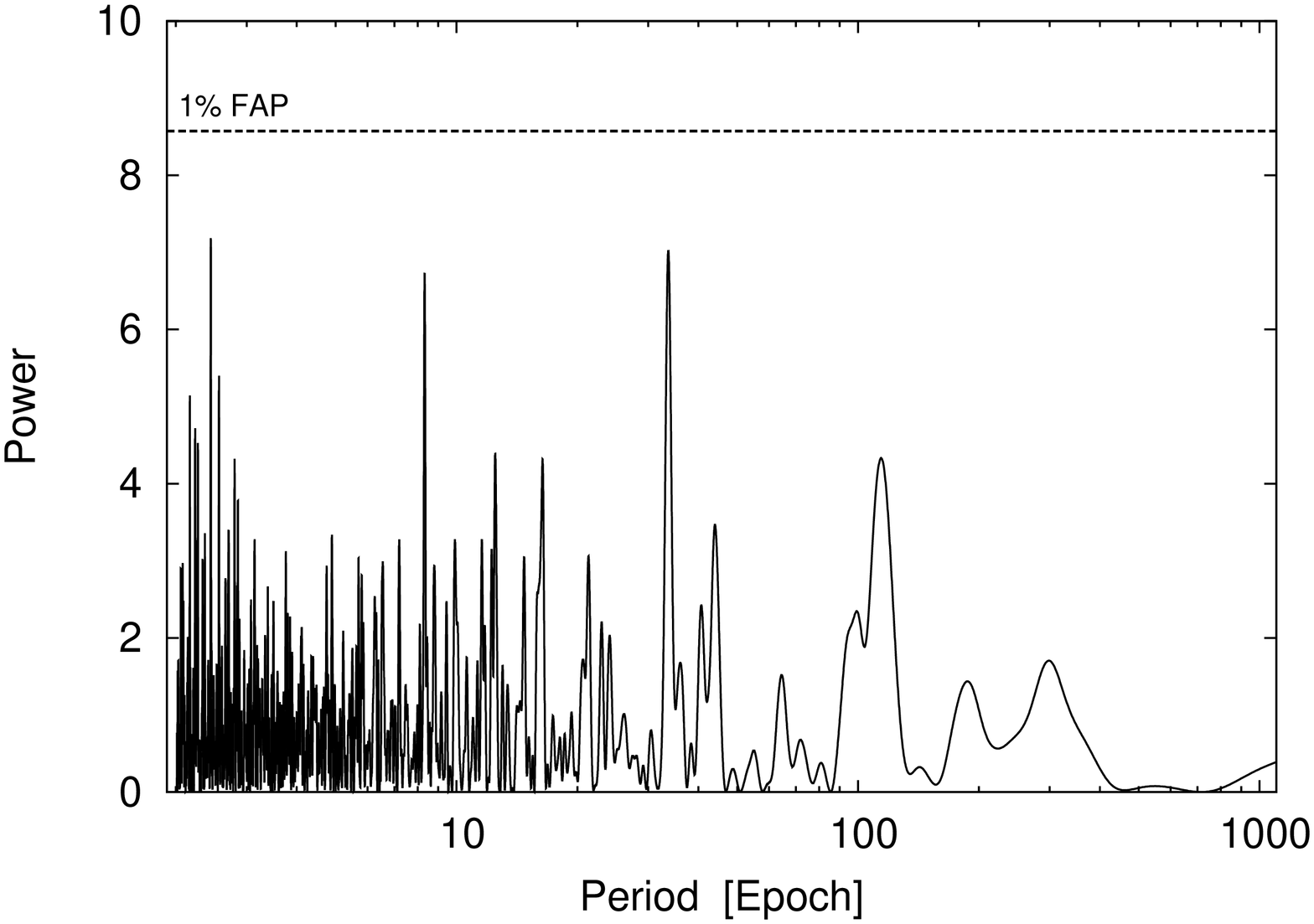}
  \caption{Lomb--Scargle periodogram for the O--C values of all transits. The dashed vertical line shows the empirical FAP level of 1\%.}
  \label{B_R_all_Fourier}
\end{minipage}
\end{figure}

For all 479 transits we determined the mid-transit time. The times have been converted from the simple Julian Date (JD) into a barycentric Julian Date based on the barycentric dynamic time (BJD$_{\mathrm{TDB}}$) using the online converter\footnote{http://astroutils.astronomy.ohio-state.edu/time/utc2bjd.html} by \citet{2010PASP..122..935E}. The downloaded data of the \textit{Kepler} telescope (Q0--Q14) are already transformed to BJD$_{\mathrm{UTC}}$. Therefore the times only had to be corrected to BJD$_{\mathrm{TDB}}$ with BJD$_{\mathrm{TDB}}$\,=\,BJD$_{\mathrm{UTC}}$\,+\,$N$\,+\,32.184\,s, where $N$ is the number of leap seconds which have elapsed since 1961. In a recent \textit{Kepler} data release \citep[Kepler Data Release Notes 20,][]{KDRN20} all transit times have been corrected to the terrestrial dynamic time (TDB), i.e. all data from Q15 are already transformed to BJD$_{\mathrm{TDB}}$.\\ To further enlarge the observational baseline we included publicly available ground-based observation in the transit timing analysis. Altogether we found 11 additional light curves in the literature: three in \citet{2007ApJ...664.1185H}, two in \citet{2009A&A...500L..45M}, five in \citet{2010A&A...510A.107M}, and one in \citet{2010ApJ...714..462S}. \\ As already mentioned in section \ref{lc_analysis} for every light curve we used two different methods to determine the uncertainties of the transit times, a MC simulation and a residual permutations algorithm (prayer bead) to assess the importance of systematic errors. We always adopted the maximum of the MC and prayer bead errors as the final uncertainties as suggested by \citet{2008MNRAS.386.1644S} and done by other authors \citep[e.g.][]{2011A&A...528A..65M}. Note that for the space-based data, where the light curves are expected to be less affected by red noise, the prayer bead method returns smaller error estimates than the MC method (see Table~\ref{TAP_vs_JKTEBOP}) and therefore we used the MC errors as the final ones. \\ Since the observations cover a period of eight years (2006--2013) and the very small uncertainties in the obtained values of the mid-time of the \textit{Kepler} transits, we could improve the linear ephemeris using a weighted linear fit with all 490 transit times (equation \ref{Elemente_TrES2}, where $E$ denotes the epoch; rms of the fit 1.31\,min). This is consistent but 3.8, 3.7, and 8.0 times more precise than the results of \citet{2011ApJ...733...36K}, \citet{2011MNRAS.417.2166S} and \citet{2012A&A...539A..97S}, respectively.
\begin{equation}
\label{Elemente_TrES2}
\begin{array}{r@{.}lcr@{.}l}
T_{\mathrm{c}}(E)=(2453957 & 6354991 & + & E\cdot  2 & 4706133738)\,\mathrm{d} \\
\pm 0 & 0000129 &  & \pm 0 & 0000000187.
\end{array}
\end{equation}
Using only the times of the \textit{Kepler} transits for the determination of the ephemeris yields $T_{\mathrm{c}}(0)=(2453957.6354983\pm0.0000130)\,\mathrm{d}$ and $P=(2.4706133749\pm0.0000000188)\,\mathrm{d}$ which is a teensy bit less precise. The inclusion of the long baseline of ground-based observation increases the precision of the orbital period as well as the time at epoch zero by a factor of $\sim$\,1.01.  \\ Fig.~\ref{O_C_Kepler} shows the O--C diagram only for the 435 \textit{Kepler} transits. To search for a periodicity in the O--C diagram of the \textit{Kepler} transits we calculated a Lomb--Scargle periodogram \citep{1976Ap&SS..39..447L,1982ApJ...263..835S}. The result, which is given in Fig.~\ref{O_C_Kepler_Fourier}, shows no significant periodicity. The false alarm probability (FAP) of this signal was determined empirically by a bootstrap resampling method. The O--C values were randomly permuted to the original epochs and then a Lomb--Scargle periodogram was calculated. This step was repeated $10^{5}$ times. The FAP of the signal is given by the fraction of periodograms that produced higher power than the original data set. The highest peak (power of 6.6) shows a FAP of 62.8\%. Within the \textit{Kepler} data we could not detect any evidence for transit timing variations (TTVs). This result was used as a starting point for a detailed investigation of the reliability of the error bars. We assumed that all 435 \textit{Kepler} measurements lie on the zero-line in the O--C diagram. Using the transit times that were obtained from the light curve analysis with \begin{scriptsize}JKTEBOP\end{scriptsize} and a simple MC simulation the weighted linear fit resulted in a $\chi^{2}$ of 507.7 which corresponds to a reduced $\chi^{2}$ of 1.17. Repeating the analysis with the transit times from the light curve analysis with \begin{scriptsize}TAP\end{scriptsize} and MCMC the fit yields a $\chi^{2}$ of  252.6 (reduced $\chi^{2}$ of 0.58). Because the $\chi^{2}$ for the \begin{scriptsize}JKTEBOP\end{scriptsize} analysis is closer to 1 than the $\chi^{2}$ from the \begin{scriptsize}TAP\end{scriptsize} analysis it is likely that the \begin{scriptsize}TAP\end{scriptsize} error bars are overestimated. To obtain a reduced $\chi^{2}$ of 1 the \begin{scriptsize}JKTEBOP\end{scriptsize} error bars of the \textit{Kepler} data have to be rescaled by a factor of 1.08. But this factor is only true for the space-based data where red noise is less important. To account for time-correlated noise in the ground-based light curve we followed a procedure that was described by \citet{2008ApJ...683.1076W}, called the ''time-averaging'' method. For every unbinned light curve we determined the standard deviation of the residuals $\sigma_{\mathrm{1}}$ for the best fit model. Then we averaged the time series into $m$ bins of $n$ data points, determined the best fit and calculated the standard deviation of the binned residuals $\sigma_{\mathrm{n}}$. In the presence of red noise $\sigma_{\mathrm{n}}$ differs from the expected value in the case of Gaussian noise by a factor $\beta$:
\begin{equation}
\beta=\frac{\sigma_{n}}{\sigma_{\mathrm{1}}}\sqrt{\frac{n(m-1)}{m}}.
\end{equation}
For this analysis we used several time bins up to 15\,min. For every light curve $\beta$ was determined as the median of these factors for the different time bins. To test if the space-base light curves are affected by red-noise we calculated the $\beta$ factor for all 435 \textit{Kepler} light curves. This analysis yielded an average of $\beta=1.08\pm0.09$ which is in excellent agreement with the value found by rescaling the error bars to obtain a weighted linear fit with a reduced $\chi^{2}$ of 1. For the ground-based data we determined the $\beta$ factor only for complete transits with a reasonable high number of data points ($>$\,100 data points otherwise the binned light curve consists of too few points for the fitting). The $\beta$ factors of the remaining 28 light curves showed a large scatter with $\beta_{\mathrm{min}}=1.00$ and $\beta_{\mathrm{max}}=1.54$ with an average of $\beta=1.25\pm0.14$. Instead of repeating all the analysis we rescaled the error bars with the factor of 1.25 (maximum of the average $\beta$ factors of ground- and space-based light curves) which seem to be resonable to take into account the quality of the ground-based light curves where conservative uncertainties are desirable. \\  For the final analysis of the O--C, the transit times of \textit{Kepler} and our ground-based data were combined. The resulting O--C diagram can be found in Fig.~\ref{O_C_all}. All transit times and O--C values are summarised in Table \ref{Transit_Times_all} (a full version of this table is available in the online version of this article). Despite the relatively low precision of some transits especially of partial transits (shown as open symbols with dashed error bars in Fig.~\ref{O_C_all}), the transit times are within the error bars largely consistent with the updated ephemeris (reduced $\chi^{2}$ of 1.05; 76.1\% of the measurements within 1$\sigma$, 96.5\% within 2$\sigma$, and 99.8\% within 3$\sigma$). Especially, one of our latest ground-based data points, obtained with the Calar Alto 2.2-m telescope (epoch 739, filled circle in Fig.~\ref{O_C_all}), has a small error bar of $\pm$\,52\,s and is fully consistent with the \textit{Kepler} data. Only few data points of the ground-based observation deviate significantly from the zero-line. Especially the transit from 2007 May 3 (epoch 108) is off by 15\,min. It seems that these points suffer from additional uncorrelated noise. At epoch 108, for example, we used a different observing mode (15 science images and three darks alternately without autoguiding). This resulted in jumps in the light curve that influenced the transit time. In most cases where we observed multiple transits with different instruments at the same epoch (i.e. epochs 163, 446, 448, 463, 622, 682, 684, 758) the transit-times are consistent with each other within the error bars. The only exception is the partial transit of epoch 737. The STK data point is consistent with the \textit{Kepler} data but the CTK-II point differs slightly which is a result of the data point gap in the ingress due to technical problems. One data point from \citet{2009A&A...500L..45M} deviates from the zero-line by 11$\sigma$ (marked by a circle in Fig.~\ref{O_C_all}). But since only the fully reduced light curve could be downloaded the reasons for this deviation could not be investigated. To test the accuracy of the calculated ephemeris we removed all transit times that differ more than 2\,$\sigma$ from the zero-line and did a recalculation with the remaining 472 data points. We found $T_{\mathrm{c}}(0)=(2453957.6354987\pm0.0000132)\,\mathrm{d}$ and $P=(2.4706133757\pm0.0000000190)\,\mathrm{d}$ which is consistent but slightly less precise than the values given in equation~(\ref{Elemente_TrES2}). The determination of the ephemeris seems to be robust against the outliers. \\ Including all transit light curves in the Lomb-Scargle calculations again resulted in no significant periodicity as shown in the periodogram in  Fig.~\ref{B_R_all_Fourier}. The highest peak (power of 7.2) still shows a FAP of 11.0\%. The transit times support variations neither on long (of the order of years, $f<0.01$) nor on short (of the order of few epochs, $f>0.1$) time-scales. Thus, we again confirm the conclusions of \citet{2011ApJ...733...36K} and \citet{2012A&A...539A..97S} with our larger data set.

\begin{table}
\centering
\caption{Transit times for all observed transits including the 435 \textit{Kepler}-transits and the publicly available transits [$^{(a)}$ Holman et al. 2007, $^{(b)}$ Mislis \& Schmitt 2009, $^{(c)}$ Mislis et al 2010, $^{(d)}$ Scuderi et al. 2010]. The O--C was calculated with the ephemeris given in equation (\ref{Elemente_TrES2}). If multiple transits per epoch are available the little numbers mark the following instruments (description in Table~\ref{CCD_Kameras}): (1) CTK, (2) RTK, (3) STK, (4) CTK-II, (5) MONICA, (6) ST6, (7) G2-1600, (8) Trebur, (9) \textit{Kepler}, (10) CAFOS, (11) Toru\'{n}. Table 13 is presented in its entiretly as online-only material. A portion is shown here for guidance regarding its form and content. The uncertainties of the mid-transit times were rescaled with a factor of 1.25 (see text).}
\label{Transit_Times_all}
\begin{tabular}{ccc}
\hline\hline
Epoch & $T_{\mathrm{c}}$ [BJD$_{\mathrm{TDB}}$] & O--C [min] \\ 
\hline
13$^{(a)}$	& $	2453989.75359	\pm	0.00023	$ & $	0.17	\pm	0.34	$	\\
15$^{(a)}$	& $	2453994.69473	\pm	0.00028	$ & $	0.04	\pm	0.41	$	\\
34$^{(a)}$	& $	2454041.63675	\pm	0.00018	$ & $	0.58	\pm	0.26	$	\\
87	& $	2454172.57798	\pm	0.00359	$ & $	-1.26	\pm	5.17	$	\\
108	& $	2454224.47213	\pm	0.00528	$ & $	14.96	\pm	7.61	$	\\
138	& $	2454298.57707	\pm	0.00306	$ & $	-4.42	\pm	4.41	$	\\
142	& $	2454308.46673	\pm	0.00205	$ & $	5.96	\pm	2.95	$	\\
163$^{(1)}$	& $	2454360.34788	\pm	0.00382	$ & $	3.47	\pm	5.51	$	\\
163$^{(6)}$	& $	2454360.34035	\pm	0.00918	$ & $	-7.38	\pm	13.22	$	\\
163$^{(5)}$	& $	2454360.34570	\pm	0.00127	$ & $	0.33	\pm	1.82	$	\\
165	& $	2454365.28565	\pm	0.00384	$ & $	-1.52	\pm	5.52	$	\\
... & ... & ... \\
\hline\hline
\end{tabular}
\end{table}

\section{Conclusions}

\begin{figure*}
  \centering
  \includegraphics[width=0.91\textwidth]{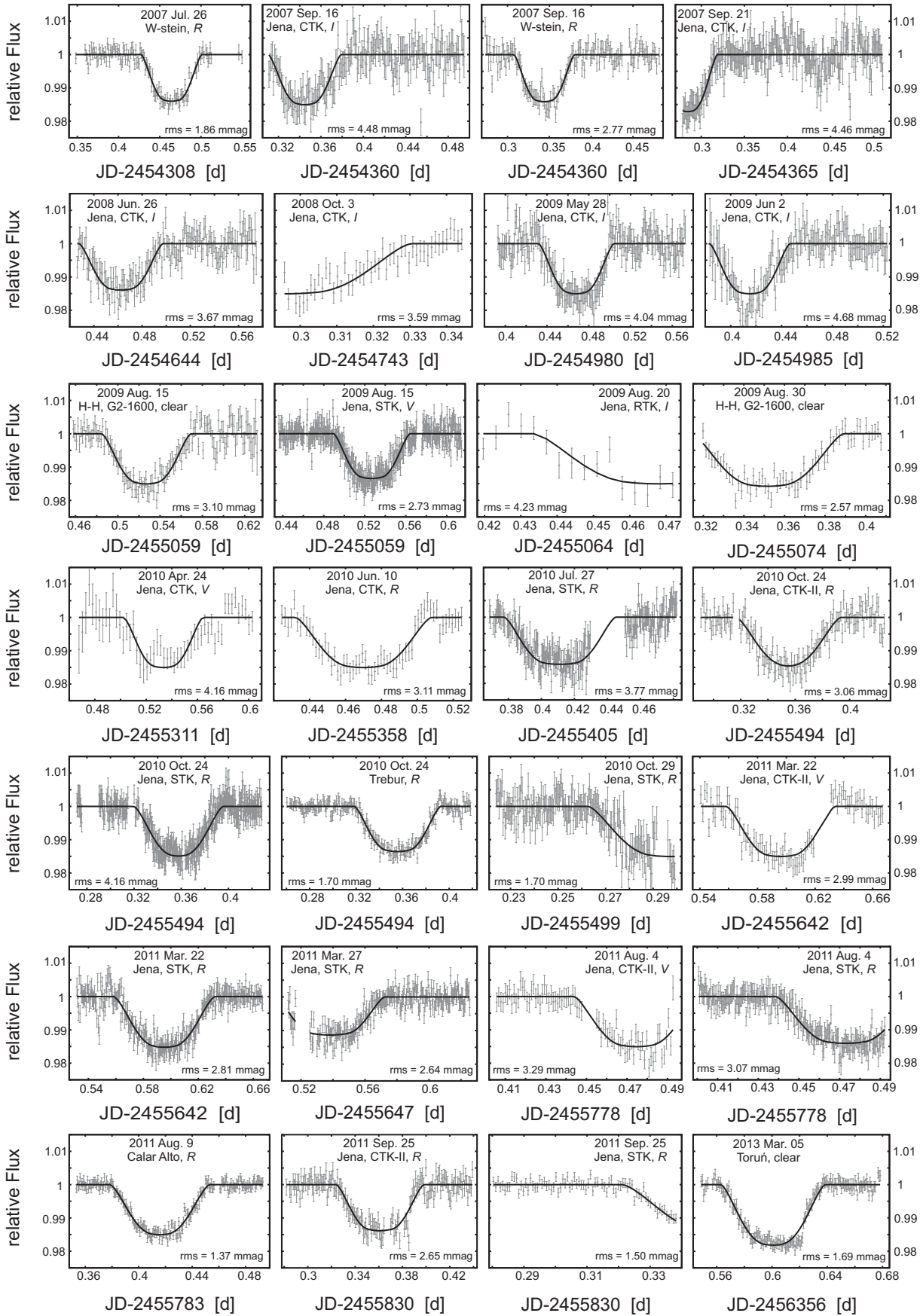}
  \caption{Light curves of TrES-2 with rms\,$<$\,4.8\,mmag. The date of observation, observatory, filter and the rms of the fit are indicated in each individual panel.}
  \label{alle_Lichtkurven_1}
\end{figure*}

\begin{figure*}
  \centering
  \includegraphics[width=0.93\textwidth]{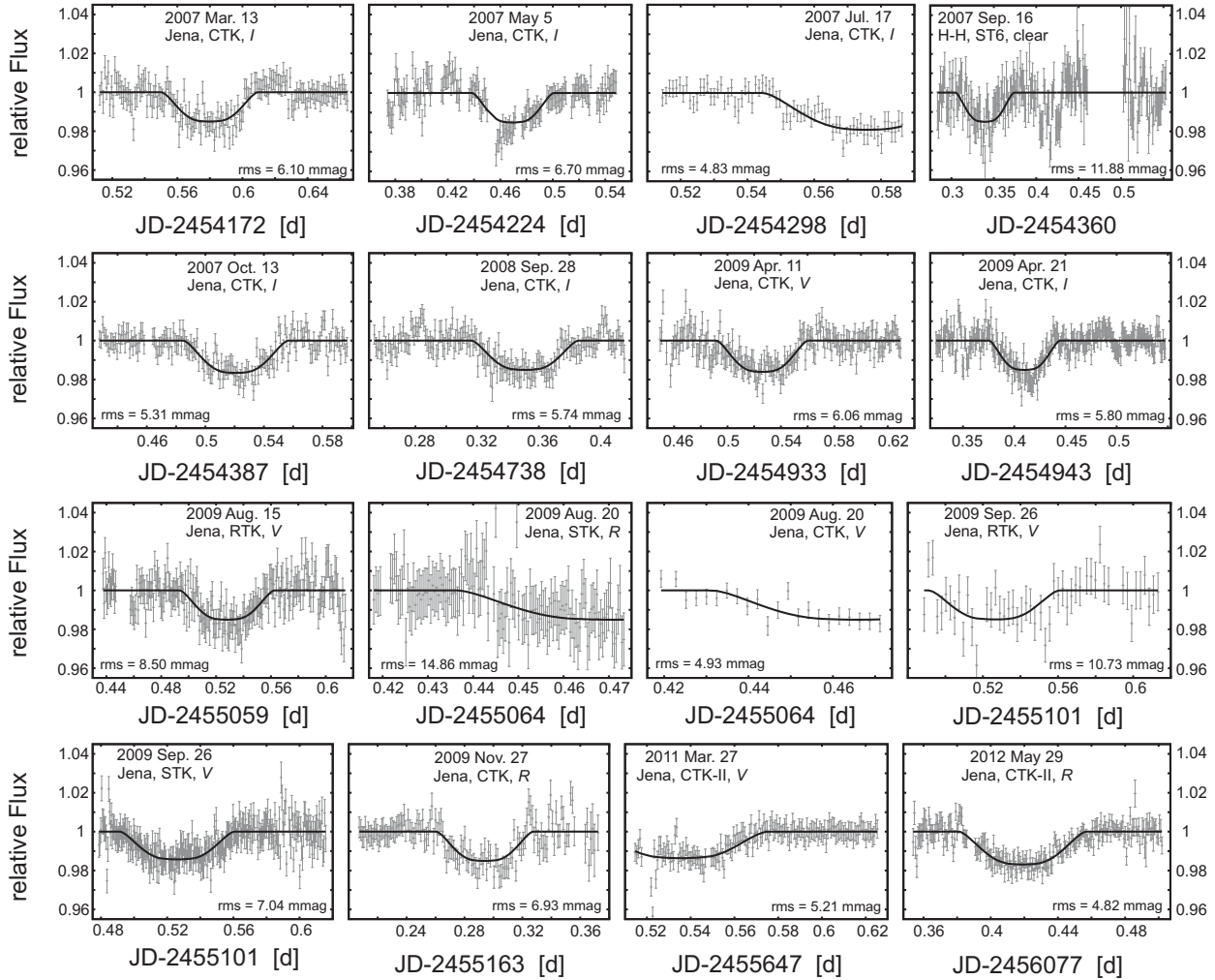}
  \caption{The same as Fig. \ref{alle_Lichtkurven_1} but for rms\,$>$\,4.8\,mmag. For a description of the observatories see Table~\ref{CCD_Kameras}. (W-stein -- Wendelstein. H-H -- Herges--Hallenberg).}
  \label{alle_Lichtkurven_2}
\end{figure*}

Subject of this work was the combined analysis of seven years of ground- and space-based observations of TrES-2 to search for transit time variations. Since very high quality of the photometry is necessary for the identification of TTVs the data reduction, photometry (for our ground-based observations) and light curve analysis were always carried out in the same manner. TrES-2 lies in the field of view of the \textit{Kepler} space telescope which offers the matchless opportunity to combine ground-based and space-based observations.\\ In this work seven years of ground-based observations were analysed (light curves are given in Figs.~\ref{alle_Lichtkurven_1} and \ref{alle_Lichtkurven_2}) along with data of 18 observation quarters (Q0--Q17) of the \textit{Kepler} space telescope. Altogether 490 individual transit mid-times were obtained and included in the calculations. The long observation period and the very small uncertainties ($\sim$\,6\,s) of the \textit{Kepler} transit times allowed a very precise redetermination of the transit ephemeris. With our much larger data set we could achieve an at least 3.7 times better precision than previous studies.\\ The insertion of all transit times in the O--C diagram showed no deviation from the zero line. Within the observation period there was neither sinusoidal variation nor recurring brief changes in transit time. Due to the continuous observation of \textit{Kepler} we especially could exclude short-term variations, i.e. changes between successive transits. However, variations on time-scales longer than the observation period are still possible. If the visual companion candidate found by \citet{2009A&A...498..567D} is indeed gravitationally bound to the TrES-2 host star (common proper motion has not yet been confirmed), it may affect the transit times (e.g. light time travel effect). To detect such changes TrES-2 would have to be followed-up over a period of several decades or even centuries.\\ The combination of our ground-based data with the \textit{Kepler} observations showed that in most cases all measurements at the same epochs are consistent within the error bars. The very high precision of the \textit{Kepler} measurements provides the opportunity to investigate further effects. With the help of a phase-folded and binned light curve we could investigate the effect of the usage of different LD laws. The light curve modelling with linear LD showed a significant deviation from the observed light curve, while the quadratic LD law reproduced the observations well.\\ Via a stepwise simultaneous light curve analysis of all 435 \textit{Kepler} light curves we could determine the parameters of the TrES-2 system up to 15 times more precisely compared to previous works. From the system parameters we calculated the physical properties of the TrES-2 system. These are within the errors completely consistent with previous studies.\\ To search for changes in the system parameters only the \textit{Kepler} transits were considered because of their unprecedented precision. We determined the parameters for each transit individually. The period of more than three years nearly continuous observations showed no statistically significant increase or decrease in the orbital inclination and the transit duration. Only the transit depth shows a slight increase which could be an indication of an increasing stellar activity. The change in spot coverage of 0.44\%  seems to be plausible compared to the solar cycle. \\ We could not confirm the change of inclination predicted in previous publications. However, within the achieved precision a slight increase cannot be excluded. \\ We found no evidence for a reoccurrence of the brightness drop 1--2 h after the transit ('dip') as the ones reported in \citet{2009AN....330..459R} and the one observed by \citet{2006ApJ...651L..61O}. Also, the \textit{Kepler} light curves showed no such event. Since the 'dip' was observed  with at least two different telescopes this could mean that it could have been caused by an accidental transit in front of a background star. Also a systematic error by short thin clouds or the existence of unusually responsive pixels on the CCD chip can not be excluded.\\ By launching several space missions like TESS \citep['Transiting Exoplanet Survey Satellite',][]{2010AAS...21545006R} from NASA's Small Explorer program, the first ESA S-class mission CHEOPS \citep['CHaracterizing ExoPlanet Satellite',][]{2013EPJWC..4703005B} or ESA's space telescope PLATO \citep['PLAnetary Transits and Oscillations of stars',][]{2011JPhCS.271a2084C} TrES-2 can still be monitored photometrically. Thus, a putative change of the system parameters or TTV can be further restricted, confirmed, or excluded.

\section*{Acknowledgements}

We would like to thank D. Keeley, H. Gilbert, and I. H\"{a}usler for participating in some of the observations at the University Observatory Jena.\\ SR, CA, RE, MK and RN would like to thank DFG for support in the Priority Programme SPP 1385 on the ``First Ten Million Years of the Solar system'' in projects NE 515/34-1 and -2, NE 515/33-1 and -2, and NE 515/35-1 and -2. MM and CG acknowledge DFG for support in program MU2695/13-1. AB would like to thank DFG for support in project NE 515 / 32-1. TE, NT, JS, RN and MMH would like to thank the DFG for support from the SFB-TR 7. NT would like to acknowledge a scholarship from the Carl-Zeiss-Stiftung. TR would like to thank DFG for support in project NE 515/36-1. CG, TOBS, and TR would like to thank DFG for support in project NE 515/30-1. CM acknowledges support from the DFG through grant SCHR665/7-1. GM, and \L{}B acknowledge the financial support from the Polish Ministry of Science and Higher Education through the Iuventus Plus grant IP2011 031971. MV, and TP would like to thank the project APVV-0158-11 and VEGA 2/0143/14. RN, GM, TP, and MV would like to thank the European Union in the Framework Programme FP6 Marie Curie Transfer of Knowledge project MTKD-CT-2006-042514 for support. RN would like to thank the German National Science Foundation (Deutsche Forschungsgemeinschaft, DFG) for general support in various projects. We would like to acknowledge financial support from the Thuringian government (B 515-07010) for the STK CCD camera used in this project.


\bibliographystyle{mn2e}
\bibliography{literatur}


\label{lastpage}

\end{document}